\documentclass[fleqn,usenatbib,useAMS]{mnras}

\usepackage{graphicx}	% Including figure files
\usepackage{amsmath}	% Advanced maths commands
\usepackage{amssymb}	% Extra maths symbols
\usepackage{multicol}        % Multi-column entries in tables
\usepackage{bm}		% Bold maths symbols, including upright Greek
\usepackage{pdflscape}	% Landscape pages

\usepackage[T1]{fontenc}
\usepackage{ae,aecompl}

\usepackage{newtxtext,newtxmath}
\newcommand{\ud}{\mathrm{d}}
\newcommand{\pd}{\partial}

\title[Lensing Cosmography]{Constraining cosmological parameters from  strong lensing  with DECIGO and B-DECIGO sources}

\author[S. Hou et al.]{Shaoqi Hou,$^{1}$%
Xi-Long Fan,$^{1}$
\thanks{E-mail: \href{mailto:xilong.fan@whu.edu.cn}{xilong.fan@whu.edu.cn}}
and Zong-Hong Zhu$^{1,2}$
\\
$^{1}$School of Physics and Technology, Wuhan University, Wuhan, Hubei 430072, China\\
$^{2}$Department of Astronomy, Beijing Normal University, Beijing 100875,  China}

\date{Accepted XXX. Received YYY; in original form ZZZ}

\pubyear{2021}

\begin{document}
\label{firstpage}
\pagerange{\pageref{firstpage}--\pageref{lastpage}}
\maketitle

\begin{abstract}
  Gravitational lensing has long been used to measure or constrain cosmology models.
  Although the lensing effect of gravitational waves has not been observed by LIGO/Virgo, it is expected that there can be a few to a few hundred lensed events to be detected by the future Japanese space-borne interferometers DECIGO and B-DECIGO, if they are running for 4 years.
  Given the predicted lensed gravitational wave events, one can estimate the constraints on the cosmological parameters via the lensing statistics and the time delay methods.
  With the lensing statistics method, the knowledge of the lens redshifts, even with the moderate uncertainties, will set the tight bound on the energy density parameter $\Omega_M$ for matter, that is, $0.288\lesssim\Omega_M\lesssim0.314$ at best.
  The constraint on the Hubble constant $H_0$ can be determined using the time delay method.
  It is found out that at $5\sigma$, $|\delta H_0|/H_0$ ranges from $3\%$ to $11\%$ for DECIGO, and B-DECIGO will give less constrained results, $8\%-15\%$.
  In this work, the uncertainties on the luminosity distance and the time delay distance are set to be $10\%$ and $20\%$, respectively.
  The improvement on measuring these distances will tighten the bounds.
\end{abstract}


%%%%%%%%%%%%%%%%%%%%%%%%%%%%%%%%%%%%%%%%%%%%%%%%%%

%%%%%%%%%%%%%%%%% BODY OF PAPER %%%%%%%%%%%%%%%%%%

\section{Introduction}

The gravitational wave (GW) was first detected about six years ago \citep{Abbott:2016blz}, verifying Einstein's prediction according to the general theory of relativity \citep{Einstein:1916cc,Einstein:1918btx}.
Since then, there have been 50 confirmed GW events collected in GWTC-1 and -2 \citep{LIGOScientific:2018mvr,Abbott:2020niy}.
Among them, GW170817 is the first neutron star-neutron star  merger event \citep{TheLIGOScientific:2017qsa}.
Most recently, the first two confident black hole-neutron star merger events GW200105 and GW200115 were reported \citep{LIGOScientific:2021qlt}.
These observations marked the new era of GW astronomy and multimessenger astrophysics.

Similar to the light, the trajectory of the GW can be bent near a massive object, resulting in the lensing effect \citep{gravlens1992,Lawrence:1971hx,Lawrence1971nc}.
In the geometric optics limit, the GW can thus travel along two or more trajectories to reach the earth, and the interferometer can detect multiple ``images''.
The lensed GW is either magnified or demagnified, so the measured luminosity distance differs from the intrinsic one.
The polarization plane changes as the GW passes by the lens, although the change can be ignored \citep{Hou:2019wdg}.
There also exist the time delays between GW signals traveling in different paths.
Under suitable conditions, multiple lensed GW signals may arrive at the interferometers simultaneously.
Since the GW generated by a binary system is highly coherent, the inference pattern can be observed, and is very useful for measuring the true luminosity distance, the mass of the lens and some cosmological parameters \citep{Hou:2019dcm}.
Generally speaking, the wavelength of the GW is much longer than that of the (visible) light, so the wave optics could be very manifest in the gravitational lensing of the GW sometimes, including ``Poisson-Arago effect" \citep{Hongsheng:2018ibg}  and the diffraction effect \citep{Nakamura:1997sw,Nakamura1999wo,Takahashi:2003ix,Liao:2019aqq}. 
However, in this work, the focus is on geometric optics, more specifically, the strong lensing effect.
This requires that the lens should be massive enough so that its curvature radius is much greater than the wavelength of the GW.

In cosmology, the Hubble constant $H_0$ is the most important parameter, quantifying the rate of the cosmological expansion.
It also determines the age and the size of our universe \citep{Weinberg:2008zzc}.
Nevertheless, the two traditional methods for measuring $H_0$ gave distinct values: $H_0=(74.03\pm1.42)\text{ km s}^{-1}\text{Mpc}^{-1}$ derived from the observation of the Cepheid variable stars, called the standard candles \citep{Riess:2019cxk}, and $H_0=(67.4\pm0.5)\text{ km s}^{-1}\text{Mpc}^{-1}$ from the data on the cosmic microwave background \citep{Aghanim:2018eyx}.
The difference is at 4.4$\sigma$, and is called the \emph{Hubble tension}.
This indicates that there is something wrong with at least one of the measurements.
One may seek for a third, independent method to measure $H_0$, which hopefully suggests certain resolution of the Hubble tension such as \citet{Dainotti:2021pqg}.
For more possible schemes to alleviate the Hubble tension, please refer to the most recent review \citet{DiValentino:2021izs}.

It is well known that GW sources can serve as standard sirens, as the luminosity distance could be obtained from the GW waveform without the help of the distance ladders \citep{Schutz:1986gp}.
Once their redshifts are determined, one can study cosmology, in particular, resolving the Hubble tension.
For binary neutron stars (BNS or NS-NS) and black hole-neutron star (BH-NS) systems, the electromagnetic counterparts exist, so their redshifts can be accurately obtained \citep{TheLIGOScientific:2017qsa,LIGOScientific:2021qlt}.
In fact, with GW170817 and its counterpart GRB 170817A \citep{Goldstein:2017mmi,Savchenko:2017ffs,Monitor:2017mdv}, it is found out that $H_0=70.0^{+12.0}_{-8.0}\text{km s}^{-1}\text{Mpc}^{-1}$ \citep{2017Natur.551...85A}.
The binary black hole (BBH or BH-BH) systems are generally not accompanied by the electromagnetic counterparts, except for the supermassive black hole binaries \citep{YanZhaoLu:2020}.
To find the redshifts of BBHs of smaller masses, one may consider to localize the host galaxies, groups or clusters of the GW sources \citep{Yu:2020vyy}.
There are other ways to measure the source redshifts without using the electromagnetic counterparts.
For example, the tidal deformation of neutron stars breaks the degeneracy between the source masses and redshift, so the GW phasing explicitly depends on the source redshift \citep{Messenger:2011gi,Messenger:2013fya}.
More generally, although the wavelength of the GW is large, it is still much smaller than the Hubble scale $H_0^{-1}$.
So the geometric optics is a good approximation to describe the GW propagation through the universe.
However,  if one considers the corrections to the geometric optics, the GW phasing also depends on the source redshift explicitly \citep{Seto:2001qf,Nishizawa:2011eq,Bonvin:2016qxr,Hou:2019jhu}.
These allow to measure the redshift by matched filtering directly.

The gravitational lensing of the GW can also be used to measure or constrain cosmological parameters, just like the lensing of light \citep{Refsdal:1964nw}.
For example, \citet{Sereno:2011ty} studied the constraining power of the lensed GW events observable by LISA with the methods of lensing statistics and time delay.
Even with one or three multi-imaged GW events to be detected in 5-year mission, one may measure $H_0$ with an accuracy of $\gtrsim 10\text{km s}^{-1}\text{Mpc}^{-1}$ and measure the energy density parameter $\Omega_M$ for matter with $\delta \Omega_M\lesssim0.08$, for instance.

In this work, we will investigate the constraints on the cosmological parameters which can be obtained using the lensed GW events observable by the future Japanese space-borne interferometer, DECi-hertz Interferometer Gravitational wave Observatory (DECIGO) \citep{Seto:2001qf,Kawamura:2011zz,Kawamura:2020pcga}, and its downscale version, B-DECIGO \citep{Sato_2017,decigo2019}.
They are sensitive to GWs at mHz to 100 Hz.
In this frequency range, there are way more binary star systems of small masses than the supermassive ones, which are the primary targets of LISA.
So in principle, DECIGO and B-DECIGO will observe more lensed GW events and produce tighter bounds on the cosmological parameters.

The gravitational lensing effect has more applications, such as constraining the speed of light \citep{Fan:2016swi,Collett:2016dey}, detecting dark matter \citep{Cutler:2009qv,Camera:2013xfa,Congedo:2018wfn,Jung:2017flg,Liao_2018},  examining the wave nature of GWs \cite{Dai:2018enj,Liao:2019aqq,Sun:2019ztn}, and probing the properties of the compact binary and galaxy populations \citep{Xu:2021bfn} and so on.
Up to now, no lensed GWs have been detected, but the advent of more sensitive detectors might make it possible soon \citep{Hannuksela:2019kle,Abbott:2021iab}.

This work is organized in the following way.
The basics of gravitational lensing of GWs is reviewed in Section~\ref{sec-regl}, where we first recall the singular isothermal sphere model in Section~\ref{sec-sis}, and then the optical depth is derived in Section~\ref{sec-tau} for predicting the lensing rates, briefly presented in Section~\ref{sec-rate}.
Section~\ref{sec-cost} introduces the basic methods for constraining the cosmological parameters: the lensing statistics in Section~\ref{sec-lss} and the time delay method in Section~\ref{sec-tdel}.
The simulation results are presented in Section~\ref{sec-sim}.
First, assuming that one can measure the source redshifts exactly, the constraints are obtained in Section~\ref{sec-ukzs}.
Second, there always exist errors in measurements, so if the errors are taken into account, the constraints become worse as shown in Section~\ref{sec-kzs}.
Finally, there is a conclusion in Section~\ref{sec-con}.
In this work, we are using the units with $G=c=1$.

\section{Review of gravitational lensing}
\label{sec-regl}

In this section, we will review some basics of the gravitational lensing effect, and the detection rates of the lensed GW events by DECIGO and B-DECIGO computed in  \citet{Piorkowska-Kurpas:2020mst}.  These are the basics for constraining the cosmological parameters \citep{Sereno:2011ty}.

\subsection{Singular isothermal sphere model}
\label{sec-sis}

According to the property of the lens, there are several different types of lensing models \citep{gravlens1992}.
For example, the simplest is the point mass model, in which a lens is just a point mass.
This is  highly idealized.
A slightly more realistic model is to treat the lens as a singular isothermal sphere (SIS).
Such kind of lens actually describes the early-type galaxy pretty well, and is characterized by the line-of-sight velocity dispersion $\sigma$.
It has a major contribution to the strong lensing probability \citep{Turner:1984ch,Moeller:2006cu}.
Of course, one may also consider more complicated models, such as the  singular isothermal ellipsoid model, but in this work, we will use the SIS model.

\begin{figure}
  \centering
  \includegraphics[width=0.45\textwidth]{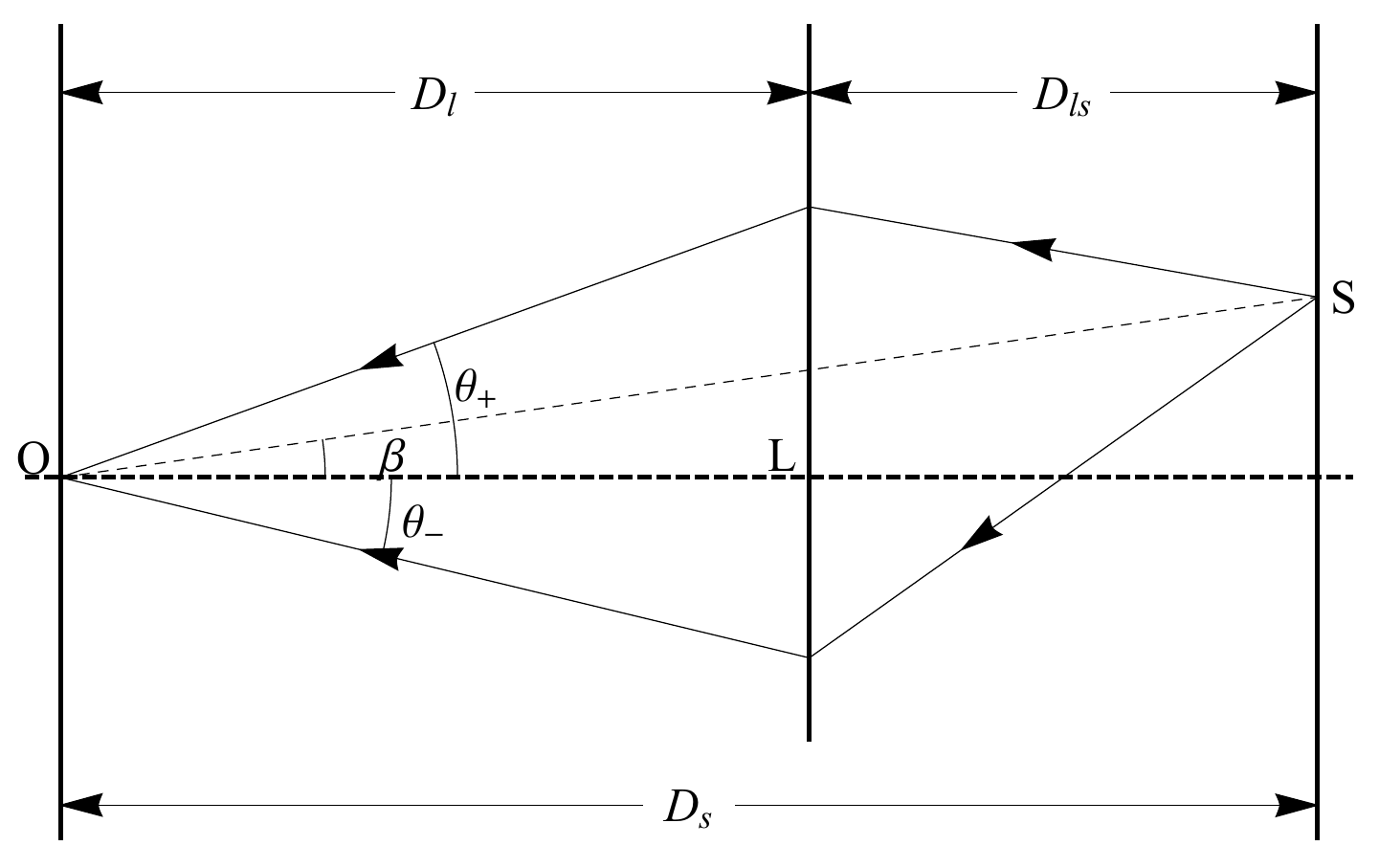}
  \caption{The geometry of a typical gravitational lensing system.
  }
  \label{fig-geo}
\end{figure}

Because of the presence of a SIS lens, the GW from a source can travel along two different trajectories to reach the earth, as shown in Fig.~\ref{fig-geo}, where two ``images'' will be detected.
We call this configuration \textit{one} multi-image event.
Two GW rays originate from the source S, travel along two trajectories, and eventually arrive at the detector at O.
Near the lens L, the two rays change their directions due to the gravitational pull of the lens.
Vertical lines represent the observer, lens and source planes, from the left to the right.
The thick dashed line is the optical axis, and the thin dashed line would be the viewing direction if the lens did not exist.
$D_\text{s}, D_\text{l},$ and $D_\text{ls}$ are angular diameter distances.
In the spatially flat Friedmann-Robertson-Walker cosmology, the angular diameter distance for an object at the redshift $z$ is \citep{Weinberg:2008zzc}
\begin{equation}
  \label{eq-daw}
  D_\text{A}(z) = \frac{1}{H_0 (1+z)} \int_0^z \frac{dz'}{E(z')},
\end{equation}
where $ E(z)=[\Omega_M (1+z)^3 + \Omega_{\Lambda}(1+z)^{3(1+w)}]^{1/2}$ is the dimensionless expansion rate with $\Omega_M$ and $\Omega_\Lambda$ the energy density parameters for the matter and the dark energy, respectively, and $w$ the dark energy equation of state.
Here, we ignore the contribution of the radiation, which is extremely small.
So $D_\text{l}=D_\text{A}(z_\text{l})$ and $D_\text{s}=D_\text{A}(z_\text{s})$, and $D_\text{ls}=D_\text{s}-D_\text{l}(1+z_\text{l})/(1+z_\text{s})$, where $z_\text{l}$ and $z_\text{s}$ are the redshifts of the lens and the source, respectively.
$\beta$ is the misalignment angle.
The GW rays form the angles $\theta_+$ and $\theta_-$ with the optical axis at the observer, satisfying \citep{gravlens1992}
\begin{equation}
  \label{eq-th-pm}
  \theta_\pm=\beta\pm\theta_\text{E},
\end{equation}
where $\theta_\text{E}=4\pi\sigma^2D_\text{ls}/D_\text{s}$ is the angular Einstein radius.
The parameter $y=\beta/\theta_\text{E}$ is often used in literature.
To have two ``images'', $y<1$ should hold.
If $y\ge1$, or there is no lens, the interferometer will observe only one ``image'', and we call this a single-image event.
When there are two ``images'', the two GW rays arrive at the detector at different times with the following time delay,
\begin{equation}\label{eq-tdv-sis}
  \Delta t  =y\Delta t_z,\quad \Delta t_z=32\pi^2\sigma^4(1+z_\text{l})D_{\Delta t},\quad D_{\Delta t}=\frac{D_\text{l}D_\text{ls}}{D_\text{s}},
\end{equation}
where the last term is called the time delay distance.
Usually, $\Delta t$ can be a few days up to approximately one month.
Finally, the amplitudes of the GW rays change with the following magnification factors \citep{gravlens1992}
\begin{equation}
  \label{eq-mus}
  \mu_\pm=\sqrt{\frac{1}{y}\pm1}.
\end{equation}
So one of the signals will be louder and the other is relatively quieter.
Because of the magnification, the measured GW signals have different signal-to-noise rations (SNRs),
\begin{equation}
  \label{eq-snrs}
  \rho_\pm=\mu_\pm \rho_0,
\end{equation}
with $\rho_0$ the intrinsic SNR.

\subsection{Optical depth}
\label{sec-tau}

In order to predict the rate of lensing, one wants to compute the optical depth $\tau$.
The differential optical depth is given by \citep{Sereno:2010dr}
\begin{equation}
  \label{eq-dtau}
  \frac{\pd^2\tau}{\pd z_\text{l}\pd\sigma}=\frac{\ud n}{\ud\sigma}s^*_\text{cr}\frac{\ud t}{\ud z_\text{l}}.
\end{equation}
Here,  $n$ is the lens number density, and $t$ is the cosmological time.
The differential number density $\ud n/\ud\sigma$ of the lens population can be described by a modified Schechter function \citep{Choi:2006qg}
\begin{equation}
  \label{eq-msf}
  \frac{\ud n}{\ud\sigma}=\frac{n_*}{\sigma_*}\frac{\gamma}{\Gamma(\alpha/\gamma)}\left( \frac{\sigma}{\sigma_*} \right)^{\alpha-1}\exp\left[ -\left( \frac{\sigma}{\sigma_*} \right)^\gamma \right],
\end{equation}
where $\Gamma(x)$ is the gamma function, and
\begin{gather*}
  n_*=8.0\times10^{-3}h^3\text{ Mpc}^{-3},\quad\sigma_*=161\pm5\text{ km/s},\\
  \alpha=2.32\pm0.10,\quad\gamma=2.67\pm0.07.
\end{gather*}
By the definition of the redshift, one has
\begin{equation}
  \label{eq-dtdz}
  \frac{\ud t}{\ud z_\text{l}}=-\frac{1}{(1+z_\text{l})H_\text{l}},\quad H_\text{l}=H(z_\text{l}).
\end{equation}
And finally, $s^*_\text{cr}$ is the cross section for lensing \citep{Hou:2020mpr}
\begin{equation}
  s^*_\text{cr}=16\pi^3\sigma^4\left( \frac{D_\text{l}D_\text{ls}}{D_\text{s}} \right)^2\left(y_\text{max}^2-\frac{2\Delta t_z}{3T_s}y_\text{max}^3\right),
\end{equation}
where $T_s(=4\text{ yrs})$ is the operation duration of DECIGO and B-DECIGO.
$y_\text{max}$ is a certain maximal value of $y$, which is determined by requiring that the two signals shall be detected by the detector, that is, $\rho_\pm\ge8$.
Therefore, one has
\begin{equation}
  \label{eq-ymax}
  y_\text{max}=\frac{1}{(8/\rho_0)^2+1}.
\end{equation}
Following \citet{Piorkowska-Kurpas:2020mst}, we set $\rho_0=8$ and thus $y_\text{max}=1/2$ to estimate the lensing rates.

With all the above information, one can integrate Eq.~\eqref{eq-dtau} once to find
\begin{equation}
  \label{eq-dtau-dzl}
  \frac{\ud\tau}{\ud z_\text{l}}=F_*\frac{D^2_{\Delta t}}{(1+z_\text{l})H_\text{l}}y_\text{max}^2\left[ 1-\frac{2\Delta t_*}{3T_s}\frac{\Gamma((8+\alpha)/\gamma)}{\Gamma((4+\alpha)/\gamma)} \right],
\end{equation}
where
\begin{gather}
  F_*=16\pi^3n_*\sigma_*^4\frac{\Gamma((4+\alpha)/\gamma)}{\Gamma(\alpha/\gamma)},\\
  \Delta t_*=32\pi^2\sigma_*^4D_\text{s}(1+z_\text{s})y_\text{max}.
\end{gather}
Integrating Eq.~\eqref{eq-dtau-dzl} again gives the optical depth,
\begin{equation}
  \label{eq-tau}
  \tau=\frac{F_*}{30}[(1+z_\text{s})D_\text{s}]^3y_\text{max}^2\left[ 1-\frac{\Delta t_*}{7T_s} \frac{\Gamma((8+\alpha)/\gamma)}{\Gamma((4+\alpha)/\gamma)}\right].
\end{equation}
Both $\tau$ and $\ud\tau/\ud z_\text{l}$ depend on the cosmological parameters through the angular diameter distances and the Hubble parameter.
Of course, $\tau$ also depends on $z_\text{s}$.

The optical depth is then used to predict the lensing rates once the merger rates are known, as briefly reviewed in the next subsection.

\subsection{Lensing rates}
\label{sec-rate}

In \citet{Piorkowska-Kurpas:2020mst}, the yearly detection rates of the lensed GW events from inspiraling double compact objects (DCOs) have been computed.
That is, the total number of the multi-image and single-image events were calculated.
DCOs include BH-BH, BH-NS and NS-NS systems.
Their intrinsic inspiral rates were calculated with the \verb+StarTrack+ population synthesis evolutionary code \citep{Dominik:2013tma}, and can be fetched from \href{https://www.syntheticuniverse.org/}{https://www.syntheticuniverse.org/}.
The intrinsic rates  were determined based on well-motivated assumptions about star formation rate, galaxy mass distribution, stellar populations, their metallicities and galaxy metallicity evolution with redshift (including ``low-end'' and ``high-end'' cases).
The binary star system evolves from zero-age main sequence to the compact binary formation after supernova (SN) explosions.
The formation of the compact object has something to do with the physics of common envelope (CE) phase of evolution and on the SN explosion mechanism.
Both of them are uncertain to some extent, so  there are four scenarios considered in \citet{Dominik:2013tma}: standard one -- based on conservative assumptions, and three of its variations -- optimistic common envelope,  delayed SN explosion and high BH kicks scenario.
Please refer to \citet{Dominik:2013tma} and references therein for more details.

The yearly detection rates depend on the inspiral rates of DCOs, and thus are related to the evolutionary scenario.
The rates for each evolutionary scenario are listed in Table~3 in \citet{Piorkowska-Kurpas:2020mst} for DECIGO and B-DECIGO.
One can find out that the BH-BH system contributes the most, ranging from a few to a few tens events per year, and it is barely possible to detect the lensed GW from NS-NS systems.
Over the 4 years' observation running, there can be one or two lensed GW events sourced by BH-NS systems.
Compared with the lensing rates reported in \citet{Sereno:2010dr} for LISA, the rates in \citet{Piorkowska-Kurpas:2020mst} are generally much larger.
So with the methods in \citet{Sereno:2011ty}, the constraints on the cosmological parameters might be stronger.

Since GW150914, there have been a few dozens of GW events detected by LIGO/Virgo collaboration and thus the merger rates were derived \citep{LIGOScientific:2018mvr,Abbott:2020gyp}.
However, the merger rates for NS-NS and BH-BH binaries still suffer from large error bars, and  the BH-NS merger rate is merely bounded from the above.
Moreover, the maximal redshift of the observed source is quite small: $z_\text{s}=0.8$.
But most of the multi-image events are sourced by binaries at redshifts around $2\sim4$ \citep{Li:2018prc,Yang:2019jhw,Piorkowska-Kurpas:2020mst,Hou:2020mpr}, then one has to reasonably extrapolate the rates to higher redshifts.
So in this work, we will not repeat the computation in \citet{Piorkowska-Kurpas:2020mst} using the observed merger rates.

\section{Cosmological tests}
\label{sec-cost}

As discussed in the previous section, the lensing effect is closely related to the cosmology model, since both $\tau$ and $\ud\tau/\ud z_\text{l}$ depend on the angular diameter distance, which is a function of some cosmological parameters via Eq.~\eqref{eq-daw}.
So the lensing effect can be used to measure these cosmological parameters, including $H_0$, $\Omega_M$ and $w$.
Since there are no observation data yet, these parameters can be constrained around the reference values $\bar H_0=70\text{ km s}^{-1}\text{Mpc}^{-1}$, $\bar\Omega_M=0.3$ and $\bar w=-1$, which were also assumed for calculating the lensing rates in \citet{Piorkowska-Kurpas:2020mst}.
The cosmology model with these reference values may be named the reference cosmology model.
For the purpose of constraining the cosmological parameters, there are two possible methods \citep{Sereno:2011ty} briefly discussed in the following subsections.

\subsection{Lensing statistics}
\label{sec-lss}

The first method is to use the following $\chi^2$ function
\begin{equation}
  \label{eq-def-chis-1}
  \chi^2_\text{ls}(\Omega_M,w)=\left[\ln P(\Omega_M,w)-\ln P(\bar\Omega_M,\bar w)\right]^2.
\end{equation}
Here, $P(\Omega_M,w)$ is a probability function, given by \citep{Kochanek:1993da,Chae:2002uf,Mitchell:2004gw,Sereno:2011ty}
\begin{equation}
  \label{eq-def-p}
  P(\Omega_M,w)=\prod_{i=1}^{N_\text{u}}(1-\tau_i)\prod_{j=1}^{N_\text{l}}p_{j},
\end{equation}
where $N_\text{u}$ is the number of single-image events and $N_\text{l}$ the number of multi-image events.
$p_j$ can be either the differential optical depth $\ud\tau_j/\ud z_{\text{l},j}$, if the redshift $z_{\text{l},j}$ of the lens for the $j$-th multi-image GW event is known, or $\tau_j$, if $z_{\text{l},j}$ is unknown.
So even if one does not identify the lens, one can still constrain $\Omega_M$ and $w$, although the knowledge of the lens redshift would greatly improve the constraints, as shown in the next section.
Since  lensing statistics are independent of $H_0$ \citep{Kochanek:1993da,Chae:2002uf}, this method does not constrain it.

In the actual observation, the source redshifts  can be obtained using the electromagnetic counterpart for BNS and BH-NS merger events, or measuring the host galaxy redshifts for BBH merger events \citep{TheLIGOScientific:2017qsa,YanZhaoLu:2020,Yu2020MNRAS.497..204Y,LIGOScientific:2021qlt}.
For the multi-image events, one could also try to identify the lenses and measure their redshifts based on the lensing analysis.

However, in the simulation, we obtain the distribution of the source redshifts for multi-image events according to the estimated differential lensing rate.
The source redshift distribution for single-image events is determined from the difference between the differential rate for the observable merger events and the differential lensing rate.
For details of how to compute these differential rates, please refer to \citet{Piorkowska-Kurpas:2020mst}.
Then, the lens redshifts can be calculated by maximizing the differential optical depth \eqref{eq-dtau-dzl} for each fixed source redshift \citep{Sereno:2011ty}.
All of the above redshifts are computed in the reference cosmology model.
Varying the cosmological parameters and using $\chi^2_\text{ls}$, one can constrain the cosmological models.

\subsection{Time delay}
\label{sec-tdel}

The time delay distance $D_{\Delta t}$ is a combination of the various angular diameter distances involved in the lensing effect; see Eq.~\eqref{eq-tdv-sis}.
So it also contains the information of the cosmology model.
Then, one can constrain the cosmological parameters \citep{Refsdal:1966MNRAS.132..101R,Saha:2006sw}.
The $\chi^2$ function in this case is defined to be
\begin{equation}
  \label{eq-chis-2}
  \chi^2_\text{td}(\Omega_M,w;h;z_\text{s})=\sum_{i=1}^{N_\text{u}}\left[ \left( \frac{D_{\Delta t}^\text{Obs.}-D_{\Delta t}}{\delta D_{\Delta t}^\text{Obs.}} \right)^2+\left( \frac{d_L^\text{Obs.}-d_L}{\delta d_L^\text{Obs.}} \right) ^2\right]_i,
\end{equation}
where $d_L$ is the luminosity distance, and the subscript $i$ at the end of the expression means to evaluate terms in the brackets for the $i$-th multi-image event.
No single-image events are used in this method.
The superscript $\text{Obs.}$ refers to the measured value, and $\delta D_{\Delta t}^\text{Obs.}$ and $\delta d_L^\text{Obs.}$ are the respective uncertainties.
With this method, one not only constrains $\Omega_M$ and $w$, but also $H_0=100h\text{ km s}^{-1}\text{Mpc}^{-1}$.
Note that this $\chi^2$ is also a function of the source redshift $z_\text{s}$.
So if $z_\text{s}$'s are unknown, they should also be treated as the model parameters \citep{Sereno:2011ty}.

In the actual measurement, to determine $D_{\Delta t}$, one may measure $\Delta t$ and infer it, based on the lens properties obtained from lensing imaging, other follow-up imaging or spectroscopic observations.
This may require the identification of the host galaxies of the sources.
The errors of the measurement include the contributions from the measurements of the time delay, the Fermat potential and the mass distribution along the line of sight \citep{Liu:2019dds}.
In our simulation, we set $\delta D_{\Delta t}/D_{\Delta t}=20\%$ \citep{Sereno:2011ty}.
By matched filtering, it is possible to obtain the luminosity distance $d_L$ \citep{Schutz:1986gp}. %, which can be used to infer the source redshift $z_\text{s}$.
%So the uncertainty of $d_L$ propagates, which causes the uncertainty of $z_\text{s}$.
Of course, for the lensed GW, the measured $d_L$ is not the actual one, as the amplitude of the GW is magnified.
If one can detect and identify the two signals of the lensed GW with the amplitudes $A_\pm$ corresponding to the two paths in Fig.~\ref{fig-geo}, one can obtain the source position \citep{Sereno:2011ty},
\begin{equation}
  \label{eq-sy}
  y=\frac{1-|A_-/A_+|}{1+|A_-/A_+|},
\end{equation}
and with Eq.~\eqref{eq-mus}, one finds the magnification factors and thus the true luminosity distance.
However,  the weak lensing effect contaminates the measured waveform, so $d_L$ is measured with the uncertainty around a few percent \citep{Holz:2005df}.
This can be partially mitigated by utilizing the shear and flexion maps to infer the convergence and thus the magnification \citep{Shapiro:2009sr,Hilbert:2010am}, so the uncertainty  can be reduced by 50\%.
We set $\delta d_L/d_L=10\%$ in the simulation.

As discussed in the previous subsection, in the simulation, we actually obtain the source and lens redshifts in the reference cosmology model.
Then, $D_{\Delta t}$ and $d_L$ are calculated in the reference cosmology model, and changing the cosmological parameters, one gets $D_{\Delta t}^\text{Obs.}$ and $d_L^\text{Obs.}$.
In this way, we determine the constraints.

\section{Simulation and constraints}
\label{sec-sim}

In this section, we will describe the procedures of the simulation, and then present the constraints on $\Omega_M,\,w$, and $h$.

\subsection{Constraints with known source redshifts}
\label{sec-ukzs}

\begin{figure*}
  \centering
  \includegraphics[width=0.9\textwidth]{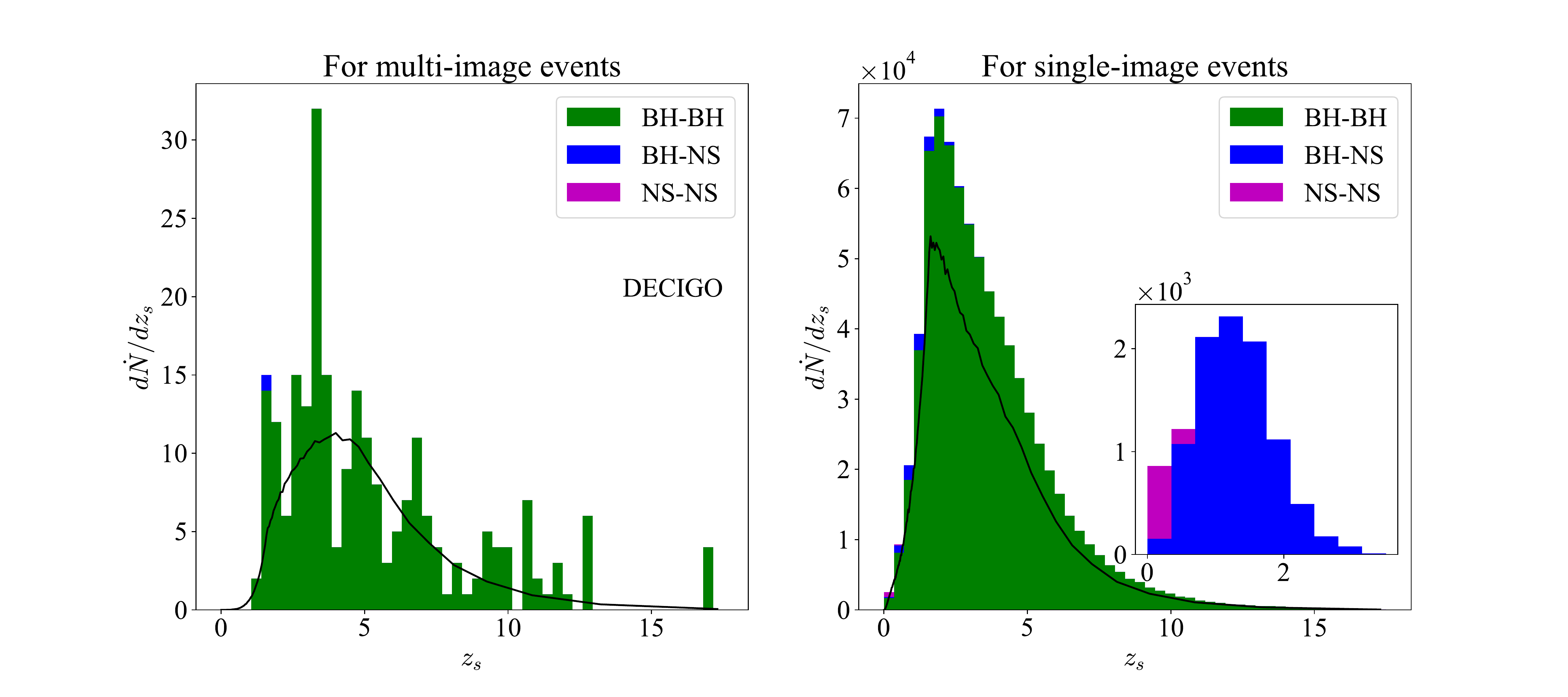}
  \caption{The \emph{yearly} differential detection rates (curves) for all the merger events and the sampled source redshift distributions (histograms) for DECIGO.
    The evolutionary scenario for DCOs is the standard one with the low-end metallicity.
  An inset is also drawn in the right panel, and only the BH-NS and NS-NS events are displayed for the better view.}
  \label{fig-rdec}
\end{figure*}

\begin{figure*}
  \centering
  \includegraphics[width=0.9\textwidth]{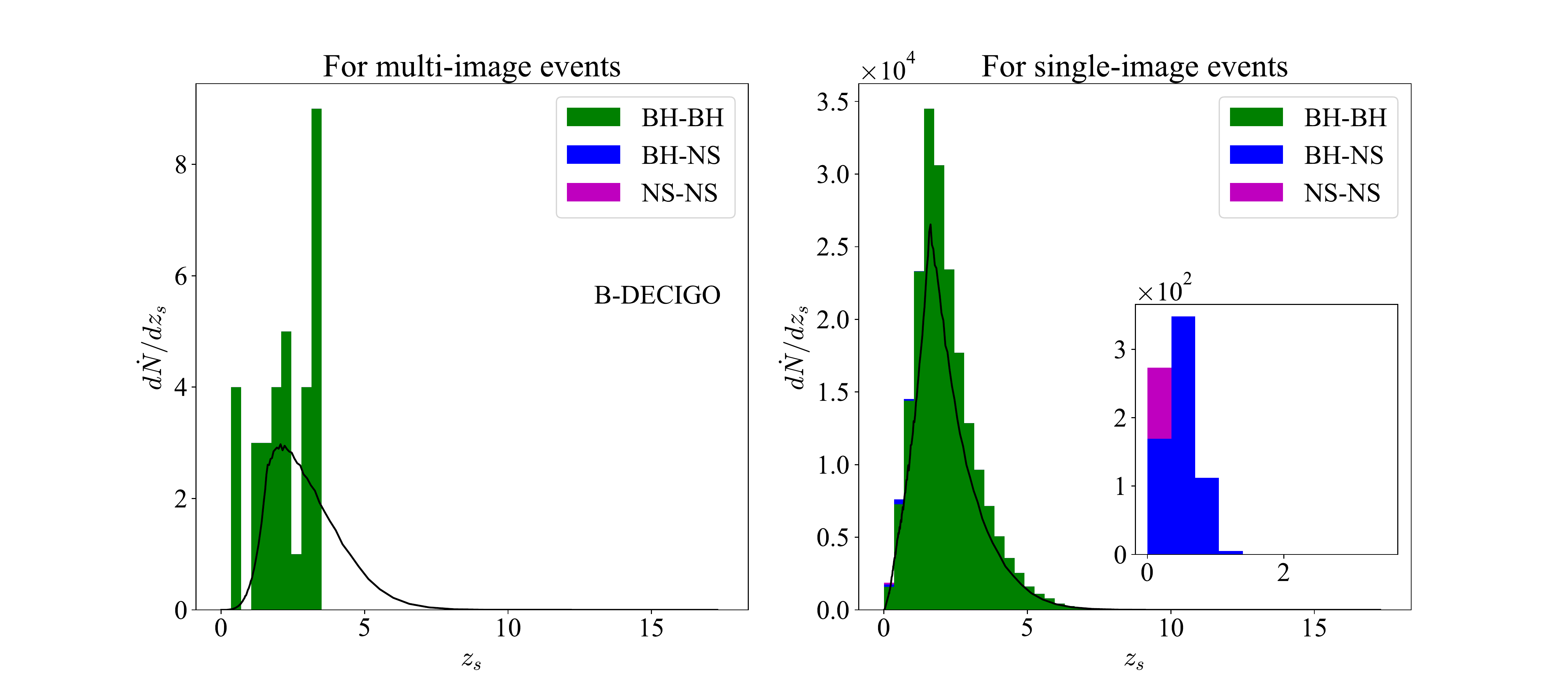}
  \caption{The \emph{yearly} differential detection rates (curves) for all the merger events and the sampled source redshift distributions (histograms) for B-DECIGO.
    The evolutionary scenario for DCOs is the standard one with the low-end metallicity.
    An inset is also drawn in the right panel, and only the BH-NS and NS-NS events are displayed for the better view.}
  \label{fig-rbdec}
\end{figure*}

In order to compute $\chi^2_\text{ls}$ and $\chi^2_\text{td}$, one calculates $\ud\tau/\ud z_\text{l}$ and  thus, the differential detection rates $\ud\dot N/\ud z_\text{s}$ of the multi-image GW events with respect to the source redshift $z_\text{s}$ \citep{Piorkowska-Kurpas:2020mst}.
With the knowledge of the total differential rates of the detectable GW events, one infers the differential rates of the single-image GW events.
These differential rates are useful for sampling the source redshifts with \verb+EMCEE+ \citep{emcee:2013} assuming that DECIGO and B-DECIGO will run for 4 years.
In Fig.~\ref{fig-rdec}, we display the \emph{yearly} differential detection rates (represented by the black curves in both panels) and the sampled source redshift distributions (represented by histograms) for DECIGO.
The evolutionary scenario for DCOs is assumed to be the standard one with the low-end metallicity.
The differential rates in both panels are actually the sums of the differential rates for three types of binary star systems.
An inset is also drawn in the right panel where only the redshift distributions the BH-NS and NS-NS events are plotted in order for people to view the distributions more clearly.
There is no inset in the left panel, since it is pretty easy to tell whether there exist BH-NS or NS-NS events or not. 
One can find out that there is no multi-image event sourced by NS-NS from the stacked histogram in the upper panel, so there are no red bars in this panel.
For B-DECIGO, the differential rates and the sampled source redshift distributions are shown in Fig.~\ref{fig-rbdec}.
In this case, all multi-image events are sourced by BH-BH.
From both figures, one notices that the histograms for the multi-image events deviate from the differential rates, because of the relatively small number of events on the order of $10\sim 10^2$.
However, for the single-image events, the deviation is minute due to the large number of events on the order of $10^5\sim10^6$.

\begin{figure*}
  \centering
  \includegraphics[width=0.9\textwidth]{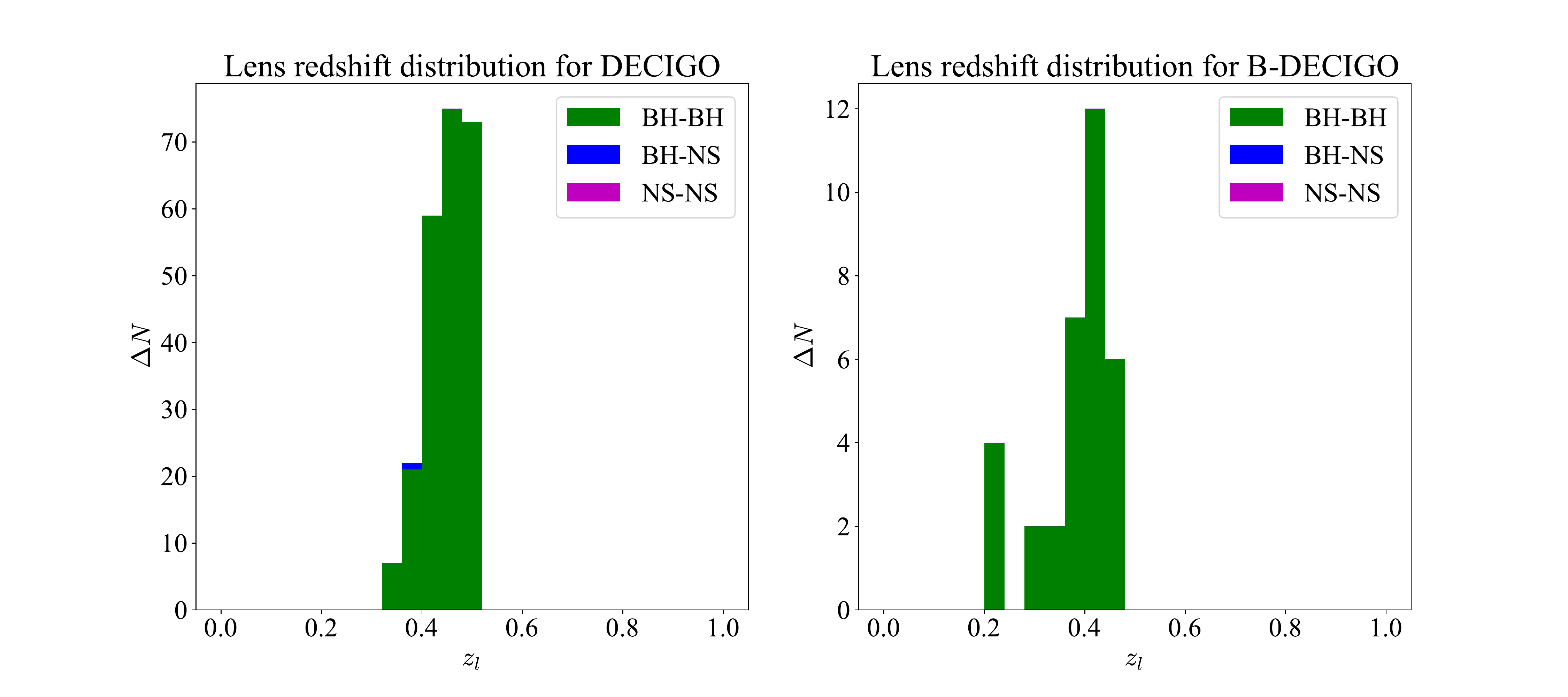}
  \caption{The lens redshift distributions for DECIGO (left panel) and B-DECIGO (right panel).
    The evolutionary scenario for DCOs is the standard one with the low-end metallicity.
  }
  \label{fig-zldis}
\end{figure*}

Once the source redshift distributions are obtained, one can determine the lens redshift distribution.
Following \citet{Sereno:2011ty}, the lens redshift for the $i$-th multi-image event is the one that maximizes $\ud\tau/\ud z_\text{l}$.
Figure~\ref{fig-zldis} displays the lens redshift distributions for DECIGO and B-DECIGO for the corresponding multi-image source redshift distributions in Figs.~\ref{fig-rdec} and \ref{fig-rbdec} (the upper panels).
Since there is no multi-image event sourced by NS-NS in the left panel of Fig.~\ref{fig-rdec}, there is no red bars in the left panel of Fig.~\ref{fig-zldis}.
Similar reason explains the absence of the blue and the red bars in the right panel of Fig.~\ref{fig-zldis}.
Although the source redshifts have wider distributions, the lens redshift distributions are confined within a narrow region ($0.2< z_\text{l}<0.6$).

\begin{figure*}
  \centering
  \includegraphics[width=0.45\textwidth]{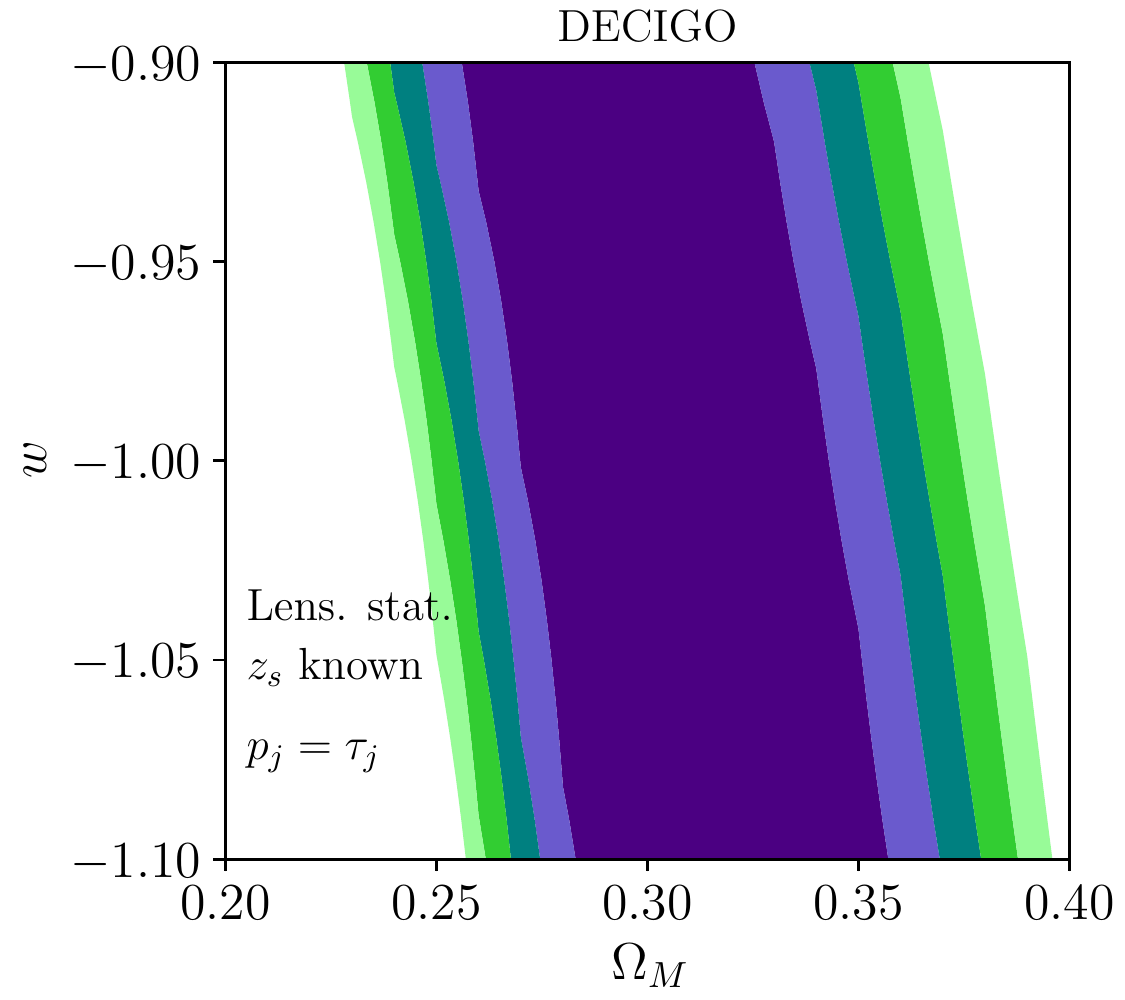}
  \includegraphics[width=0.45\textwidth]{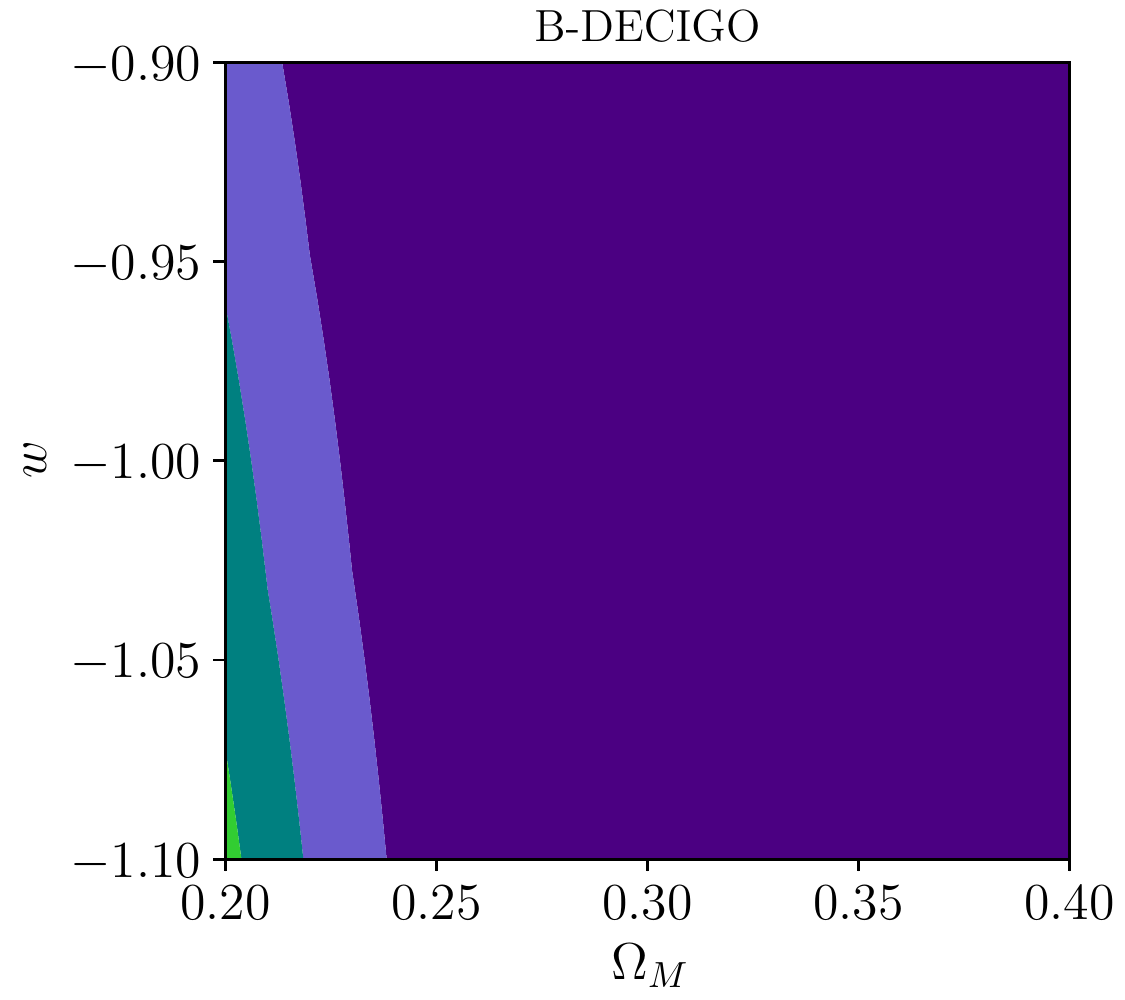}
  \caption{Constraints on $\Omega_M$ and $w$ from lensing statistics  with $p_j=\tau_j$ in Eq.~\eqref{eq-def-chis-1} for DECIGO (left panel) and B-DECIGO (right panel).
    The contours correspond to $1\sigma,\,2\sigma,\cdots,5\sigma$ confidence limit regions from inside to outside.
    The label "Lens. stat." implies that the lensing statistics method is used.}
  \label{fig-c-ls-zlu}
\end{figure*}

\begin{figure*}
  \centering
  \includegraphics[width=0.45\textwidth]{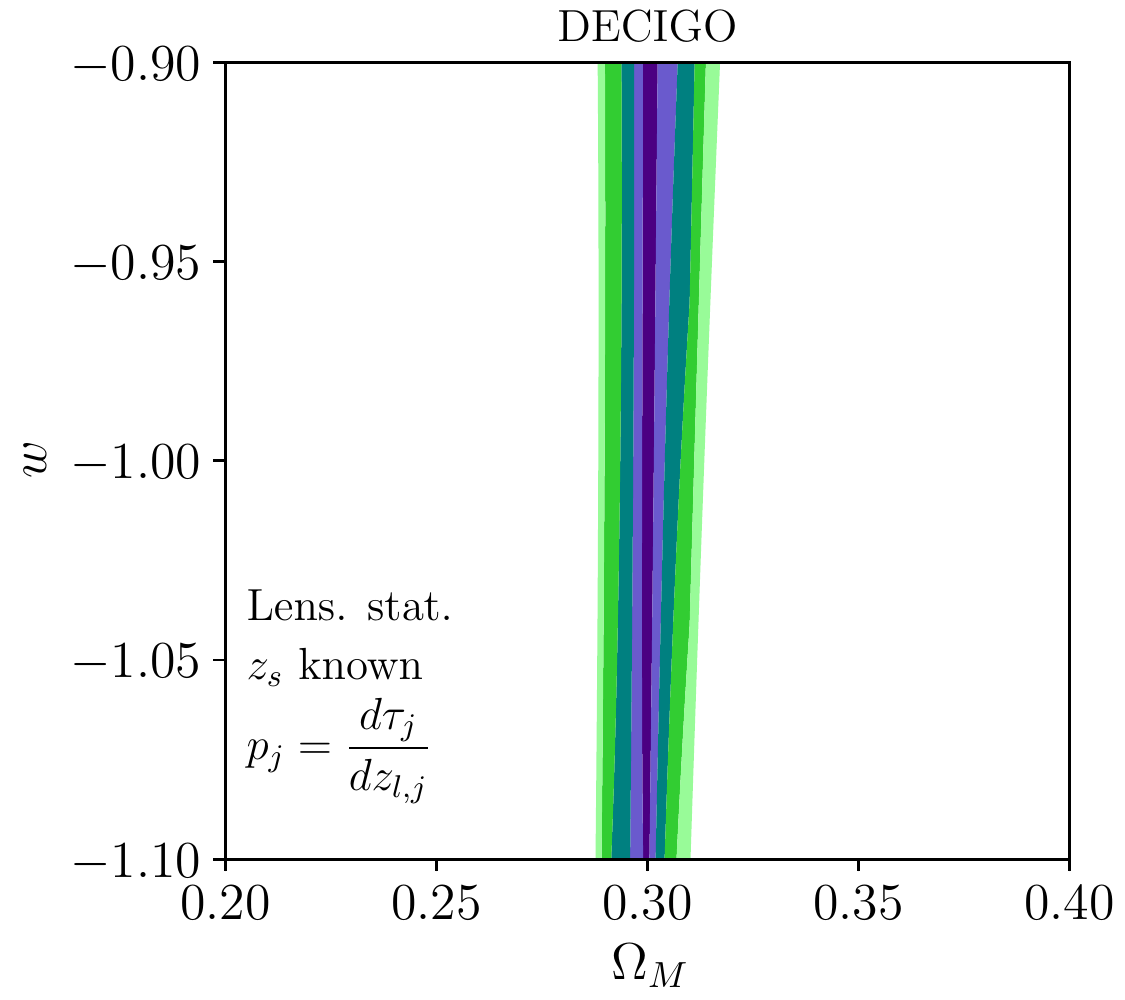}
  \includegraphics[width=0.45\textwidth]{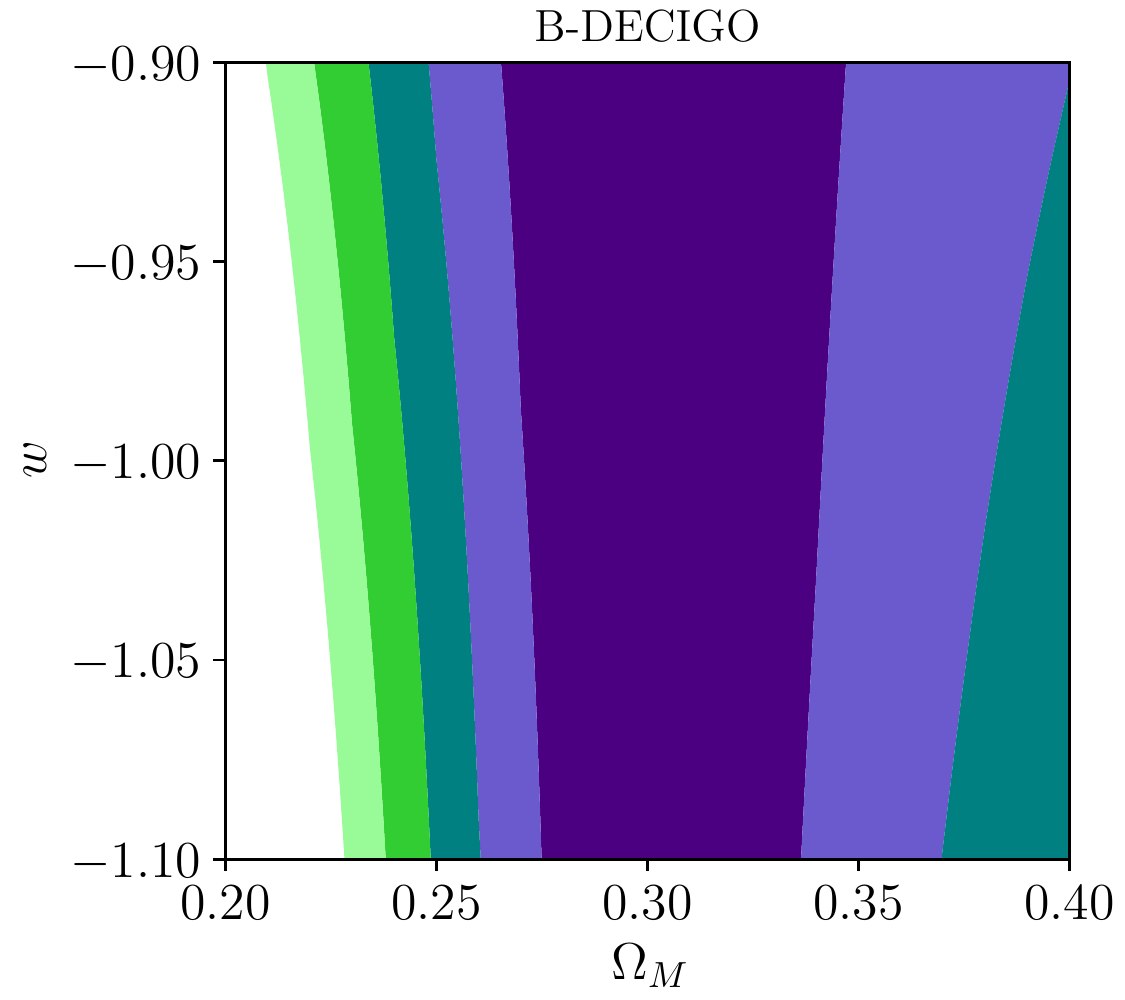}
  \caption{Constraints on $\Omega_M$ and $w$ from lensing statistics with $p_j=\ud\tau_j/\ud z_{\text{l},j}$ in Eq.~\eqref{eq-def-chis-1} for DECIGO (left panel) and B-DECIGO (right panel).}
  \label{fig-c-ls-zlk}
\end{figure*}

With these sampled redshifts, one can constrain $\Omega_M$, $w$ and $h$ using the lensing statistics and time delay methods in the following way.
First, consider the lensing statistics method.
The sampled source redshifts for the multi-image and the single-image events can be fed into Eq.~\eqref{eq-def-chis-1} to constrain $\Omega_M$ and $w$.
In this case, $p_j$ in Eq.~\eqref{eq-def-chis-1} should be the optical depth $\tau_j$, as we do not use the lens redshifts at this step, i.e., $z_\text{l}$ is assumed to be  unknown.
In our simulation, we chose $0.2\le\Omega_M\le0.4$ and $-1.1\le w\le-0.9$ with the flat priors, instead of broader ranges as used in \citet{Sereno:2011ty}.
This is because in our simulation, the DCO merger rates from \verb+StarTrack+ are used. 
These rates should depend on the cosmological parameters.
Since only the rates for the reference cosmology model are published, it is better to set the values of $\Omega_M$ and $w$ in narrower ranges in which the \verb+StarTrack+ rates may still approximately apply.
Then, we obtain the constraints as shown in Fig.~\ref{fig-c-ls-zlu}.
But if the sampled source redshifts for the single-image events and the lens redshifts (for the multi-image events, of course) are used, then $p_j$ can be $\ud\tau_j/\ud z_{\text{l},j}$.
In this case, $z_\text{l}$ is assumed to be known.
Then, one can still constrain $\Omega_M$ and $w$, given in Fig.~\ref{fig-c-ls-zlk}.
From both figures, one finds out that DECIGO  provides better constraints than B-DECIGO, since the former is more sensitive.
$\Omega_M$ is constrained within the chosen parameter region, but $w$ is unconstrained, unfortunately.
By comparing the corresponding panels in  Figs.~\ref{fig-c-ls-zlu} and \ref{fig-c-ls-zlk}, one knows that the choice $p_j=\ud\tau_j/\pd z_{\text{l},j}$ leads to tighter bound on $\Omega_M$, and in particular, the constraint on $\Omega_M$ shown in the left panel of Fig.~\ref{fig-c-ls-zlk} for DECIGO is very strong: $0.288\lesssim\Omega_M\lesssim0.313$ at $5\sigma$.

\begin{figure*}
  \centering
  \includegraphics[width=0.45\textwidth]{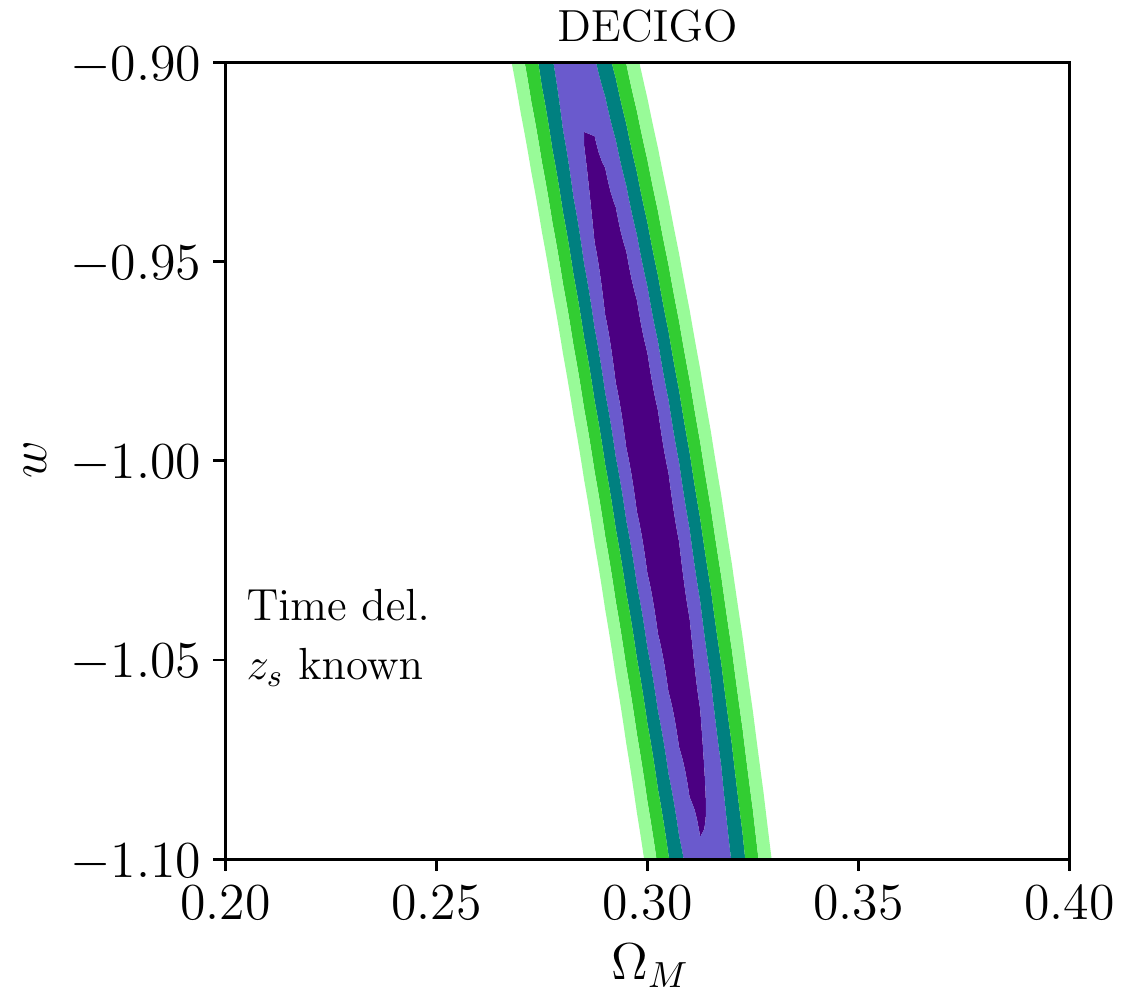}
  \includegraphics[width=0.45\textwidth]{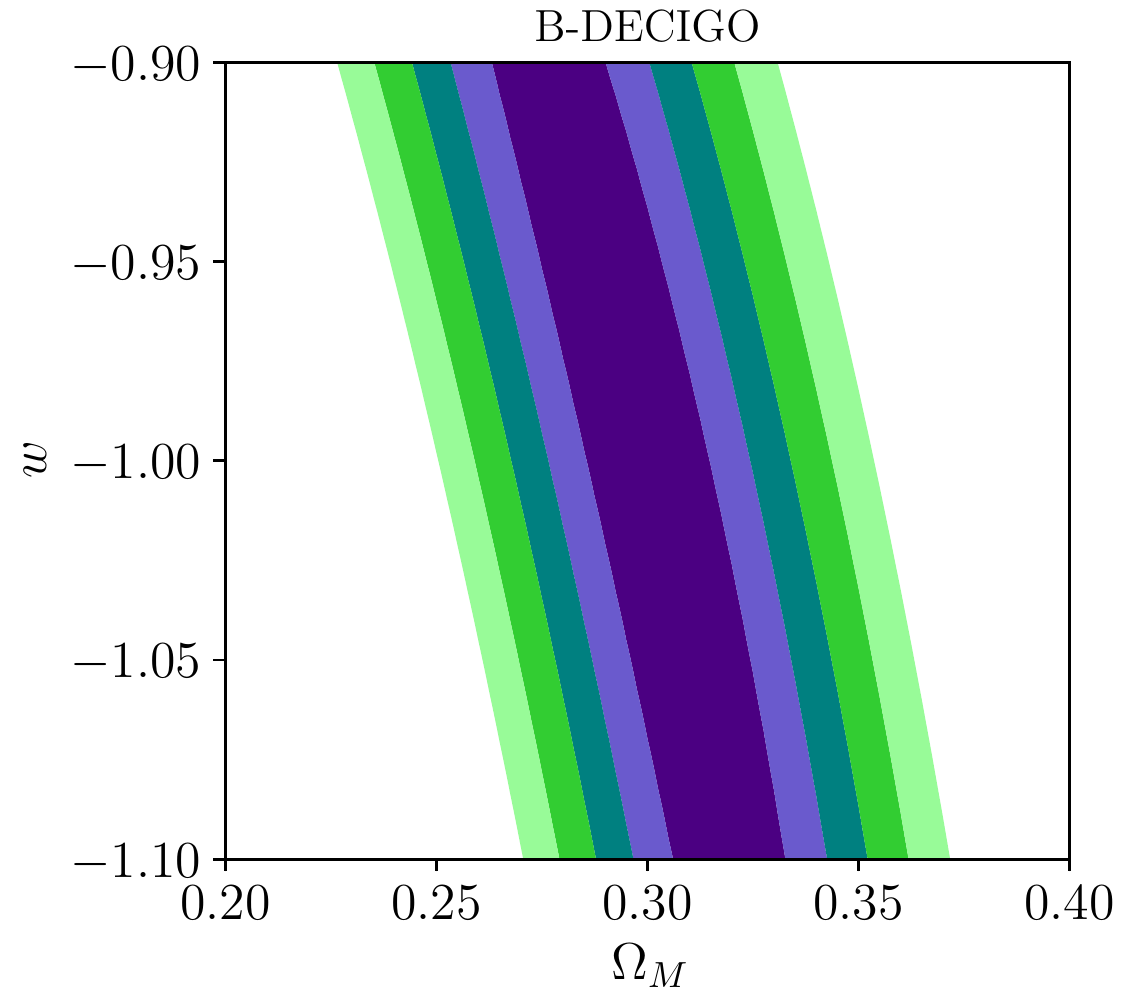}
  \caption{Constraints on $\Omega_M$ and $w$ from time delay method for DECIGO (left panel) and B-DECIGO (right panel).}
  \label{fig-c-td-nh0-k}
\end{figure*}

\begin{figure*}
  \centering
  \includegraphics[width=0.45\textwidth]{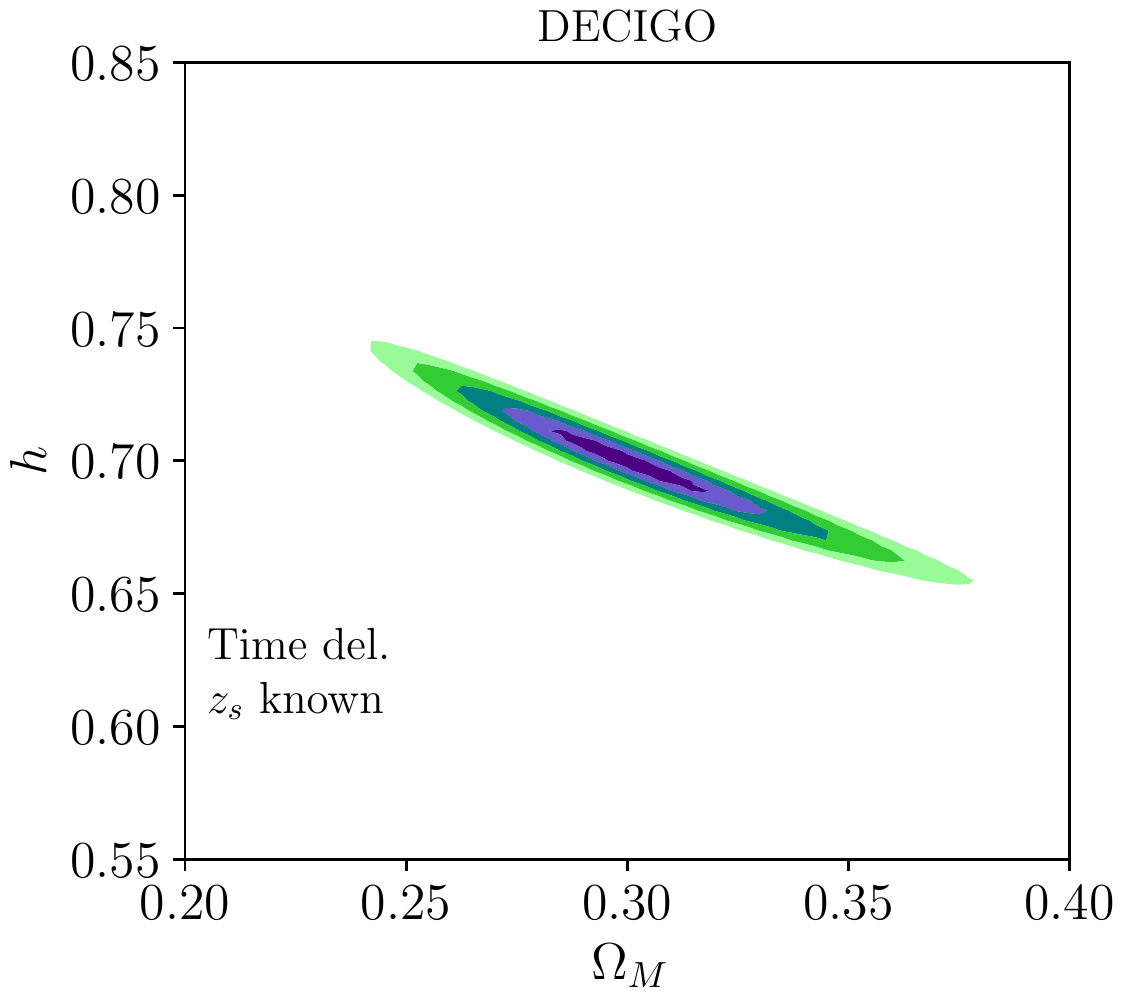}
  % the lower and upper bounds on h: [0.678,0.722]
  \includegraphics[width=0.45\textwidth]{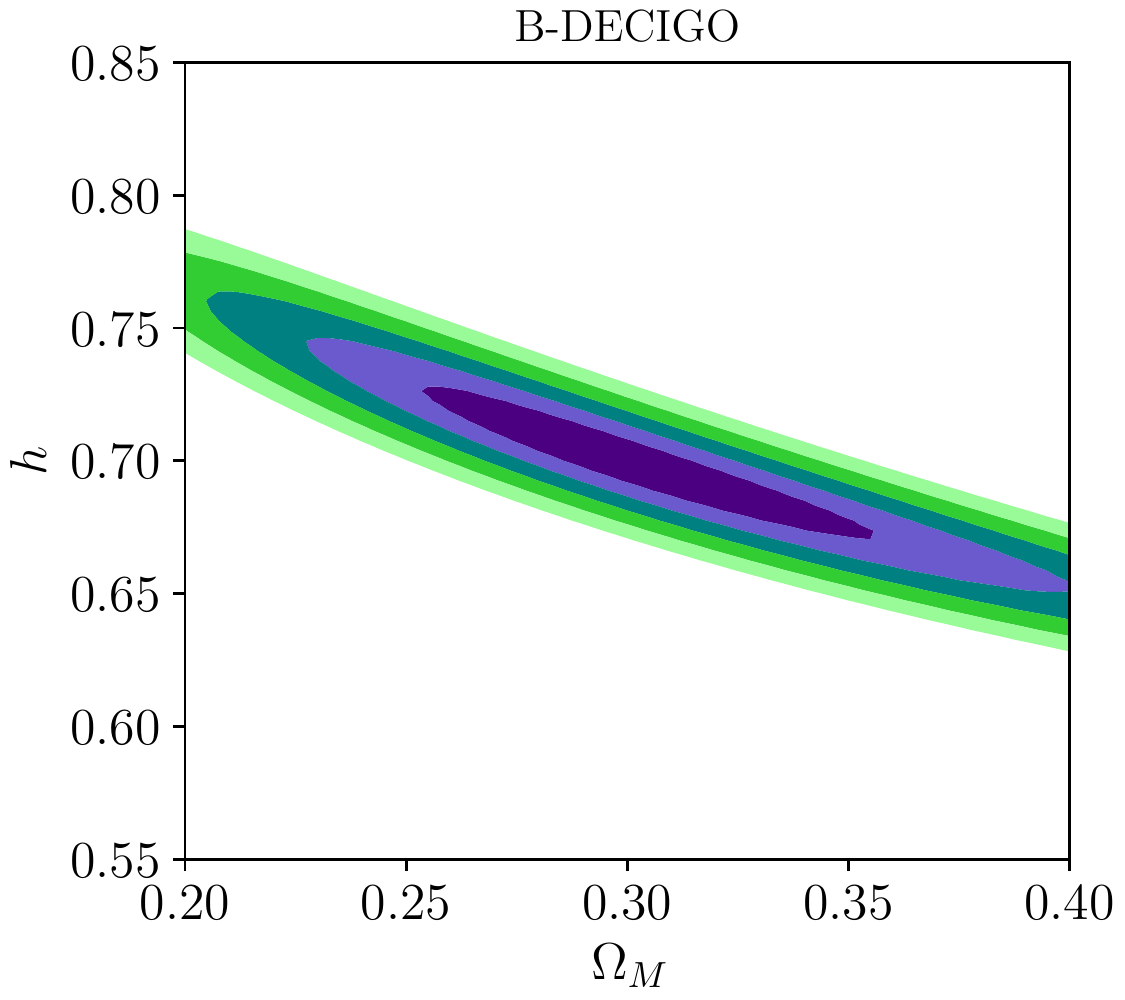}
  % the lower and upper bounds on h: [0.641,0.758]
  \caption{Constraints on $\Omega_M$ and $h$ from time delay method for DECIGO (left panel) and B-DECIGO (right panel).}
  \label{fig-c-td-h0-k}
\end{figure*}

Second, consider the time delay method.
Then, one uses the sampled source and lens redshifts for the multi-image events and Eq.~\eqref{eq-chis-2}.
With this method, one can still constrain $\Omega_M$ and $w$ at $h=0.7$.
Again, one sets $0.2\le\Omega_M\le0.4$ and $-1.1\le w\le-0.9$ with the flat priors.
The results are shown in Fig.~\ref{fig-c-td-nh0-k}.
Or, one can constrain $\Omega_M$ and $h$ with $w$ set to 0.3.
$\Omega_M$ still takes values in the range $[0.2,0.4]$ and one sets $h\in[0.6,0.8]$.
In this case, since one knows the source redshifts, the number of model parameters is 2 (i.e., $\Omega_M$ and $h$).
The constraints are displayed in Fig.~\ref{fig-c-td-h0-k}.
By comparing Fig.~\ref{fig-c-td-nh0-k} with Fig.~\ref{fig-c-ls-zlk}, the constraint on $\Omega_M$ becomes worse for DECIGO, and for B-DECIGO, the constraint does not improve very much.
From Fig.~\ref{fig-c-td-h0-k}, at $5\sigma$ level, one has $0.678\lesssim h\lesssim0.722$ by DECIGO and $0.641\lesssim h\lesssim0.758$ by B-DECIGO at $\Omega_M=0.3$.
The bounds on $h$ become stronger by a factor of at least 10 in contrast to these listed in Table~1 in \citet{Sereno:2011ty}.

\subsection{Constraints with unknown source redshifts}
\label{sec-kzs}

In the above setup, we simply use the sampled source and lens redshifts.
This equivalently means that the redshifts can be measured exactly.
However, this is not true in reality.
So in this subsection, we perform a more realistic simulation by adding uncertainties to the source and lens redshifts.

\begin{figure*}
  \centering
  \includegraphics[width=0.9\textwidth]{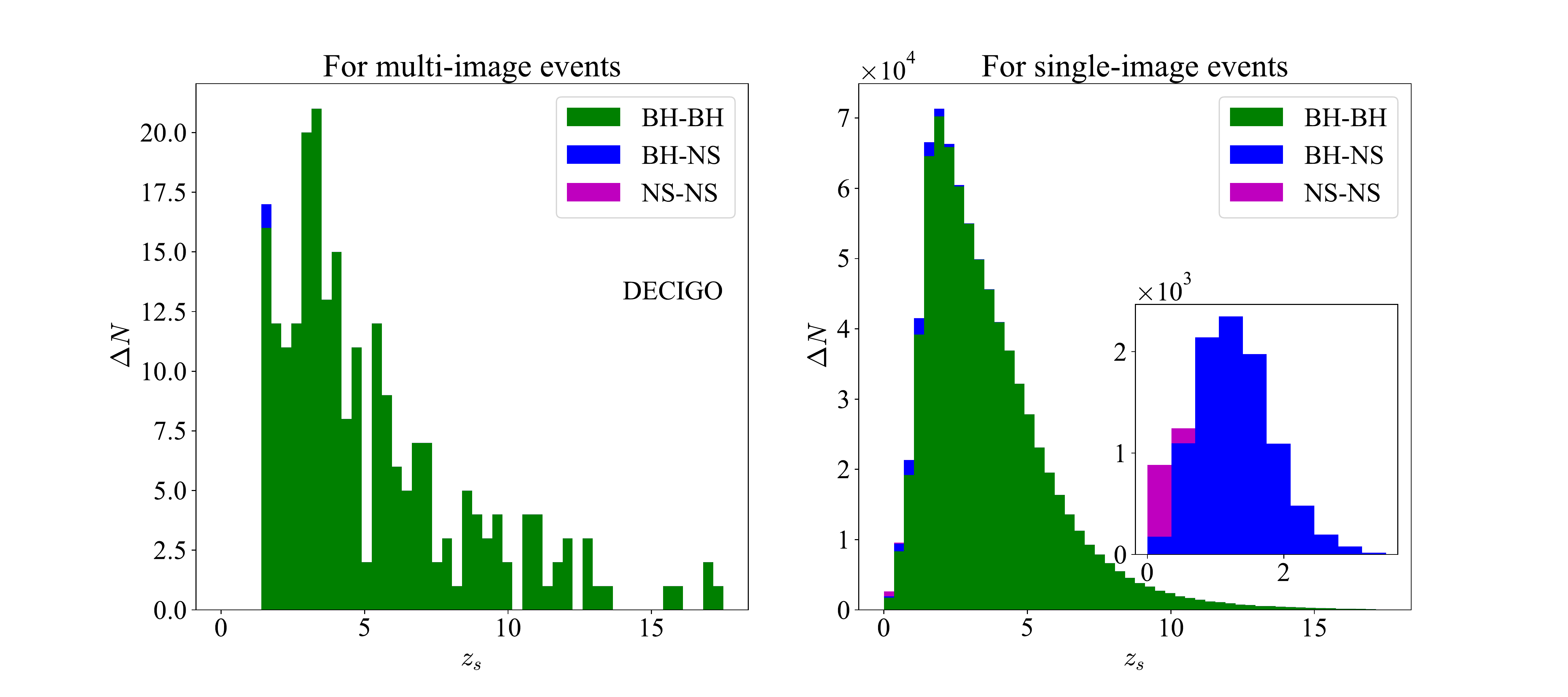}
  \caption{The sampled source redshift distributions for DECIGO assuming $\delta d_L/d_L=10\%$.
    The evolutionary scenario for DCOs is the standard one with the low-end metallicity.
 An inset is also drawn in the right panel, and only the BH-NS and NS-NS events are displayed for the better view. }
  \label{fig-rdec-u}
\end{figure*}

\begin{figure*}
  \centering
  \includegraphics[width=0.9\textwidth]{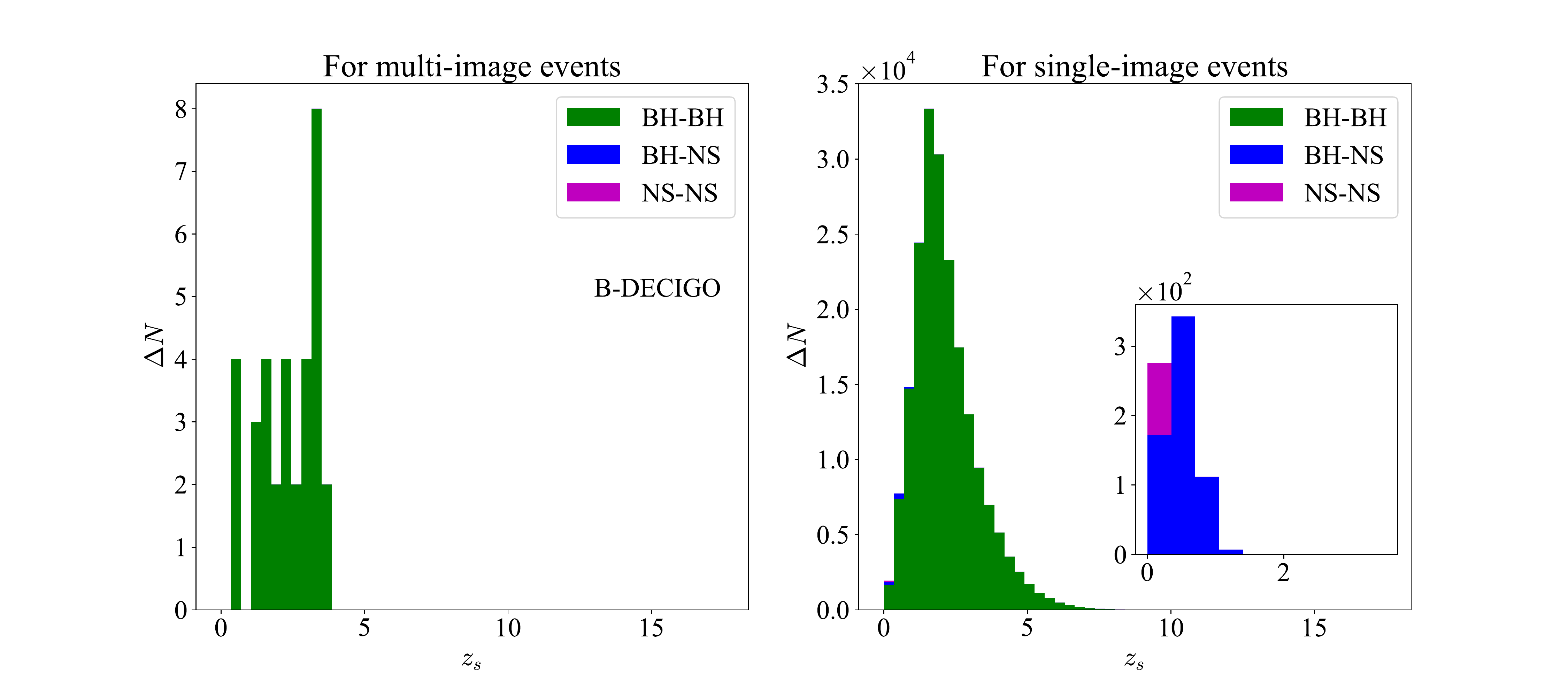}
  \caption{The sampled source redshift distributions for B-DECIGO assuming $\delta d_L/d_L=10\%$.
    The evolutionary scenario for DCOs is the standard one with the low-end metallicity. An inset is also drawn in the right panel, and only the BH-NS and NS-NS events are displayed for the better view.}
  \label{fig-rbdec-u}
\end{figure*}

%\cite{Nakamura:2016hna} discussed the uncertainty in measuring the luminosity distance, but it depends on the source redshift and quite huge at the large redshifts $z\in(10,30)$.
In our simulation, we set the uncertainty of $d_L$ to be $\delta d_L/d_L=10\%$ \citep{Sereno:2011ty}.
Randomly change $d_L$, and thus, obtain the ``measured'' source redshift distributions in Figs.~\ref{fig-rdec-u} and \ref{fig-rbdec-u} for DECIGO and B-DECIGO, respectively.

\begin{figure*}
  \centering
  \includegraphics[width=0.9\textwidth]{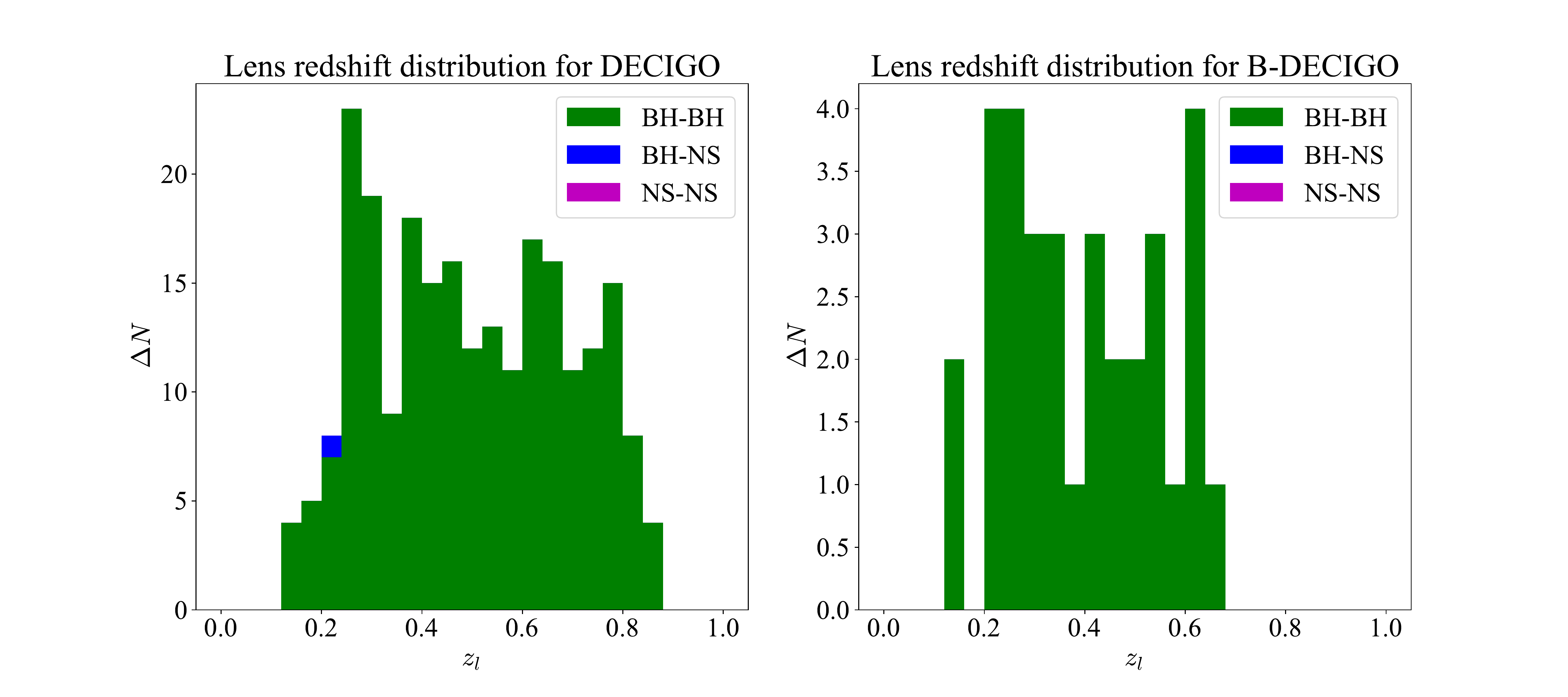}
  \caption{The lens redshift distributions for DECIGO and B-DECIGO assuming $\delta d_L/d_L=10\%$ and $\delta D_{\Delta t}/D_{\Delta t}=20\%$.
  The evolutionary scenario for DCOs is the standard one with the low-end metallicity.
  }
  \label{fig-zldis-u}
\end{figure*}

Similarly, one can determine the ``measured'' lens redshift distribution.
Instead of maximizing $\ud\tau/\ud z_\text{l}$ as did in the last subsection, the ``measured'' $z_\text{l}$ is obtained from the time delay distance $D_{\Delta t}$ with a certain measurement error $\delta D_{\Delta t}$, given the ``measured'' source redshift $z_\text{s}$.
We set the uncertainty of $D_{\Delta t}$ to $\delta D_{\Delta t}/D_{\Delta t}=20\%$ \citep{Sereno:2011ty}, and the modified lens redshift distributions are shown in Fig.~\ref{fig-zldis-u}.
These distributions are broader than the corresponding ones in Fig.~\ref{fig-zldis}.

\begin{figure*}
  \centering
  \includegraphics[width=0.45\textwidth]{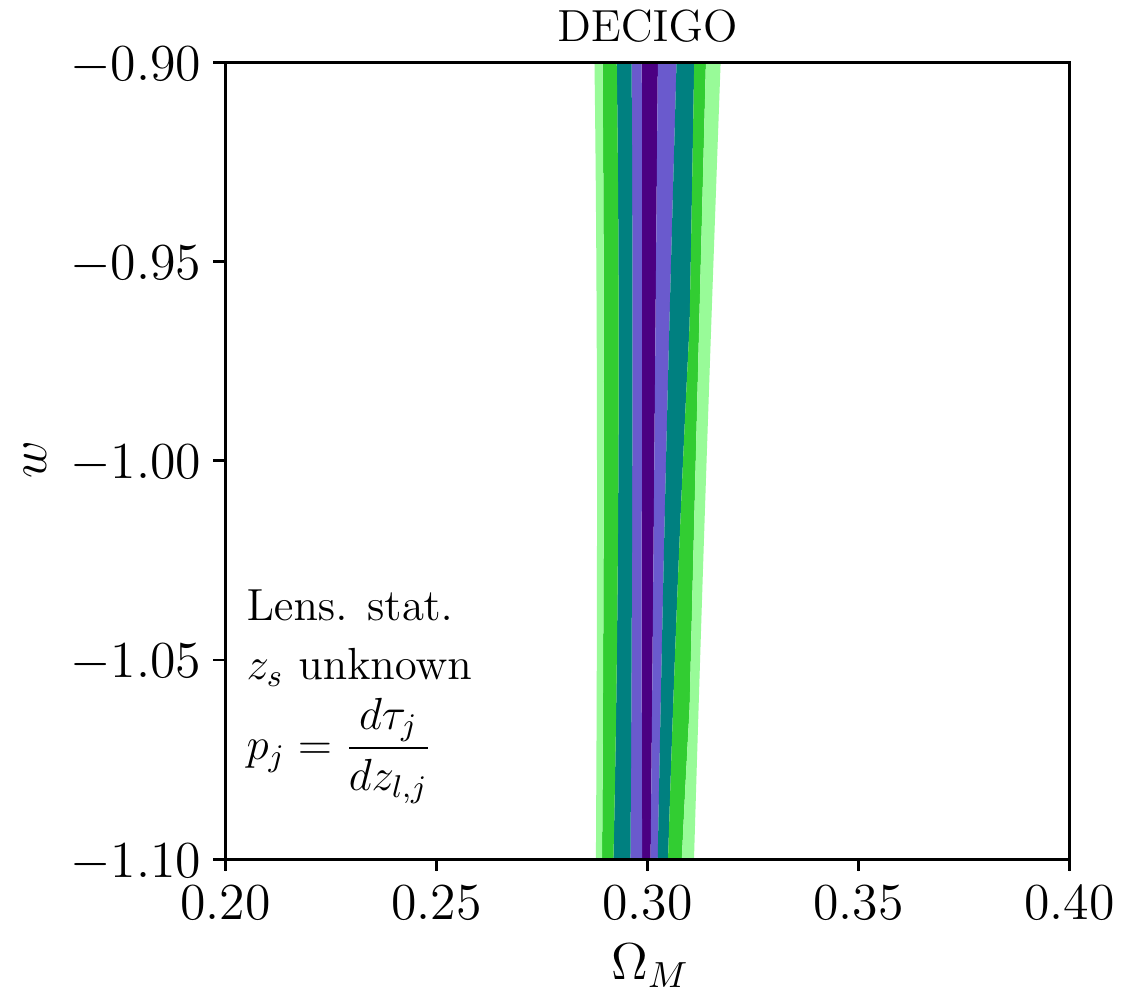}
  \includegraphics[width=0.45\textwidth]{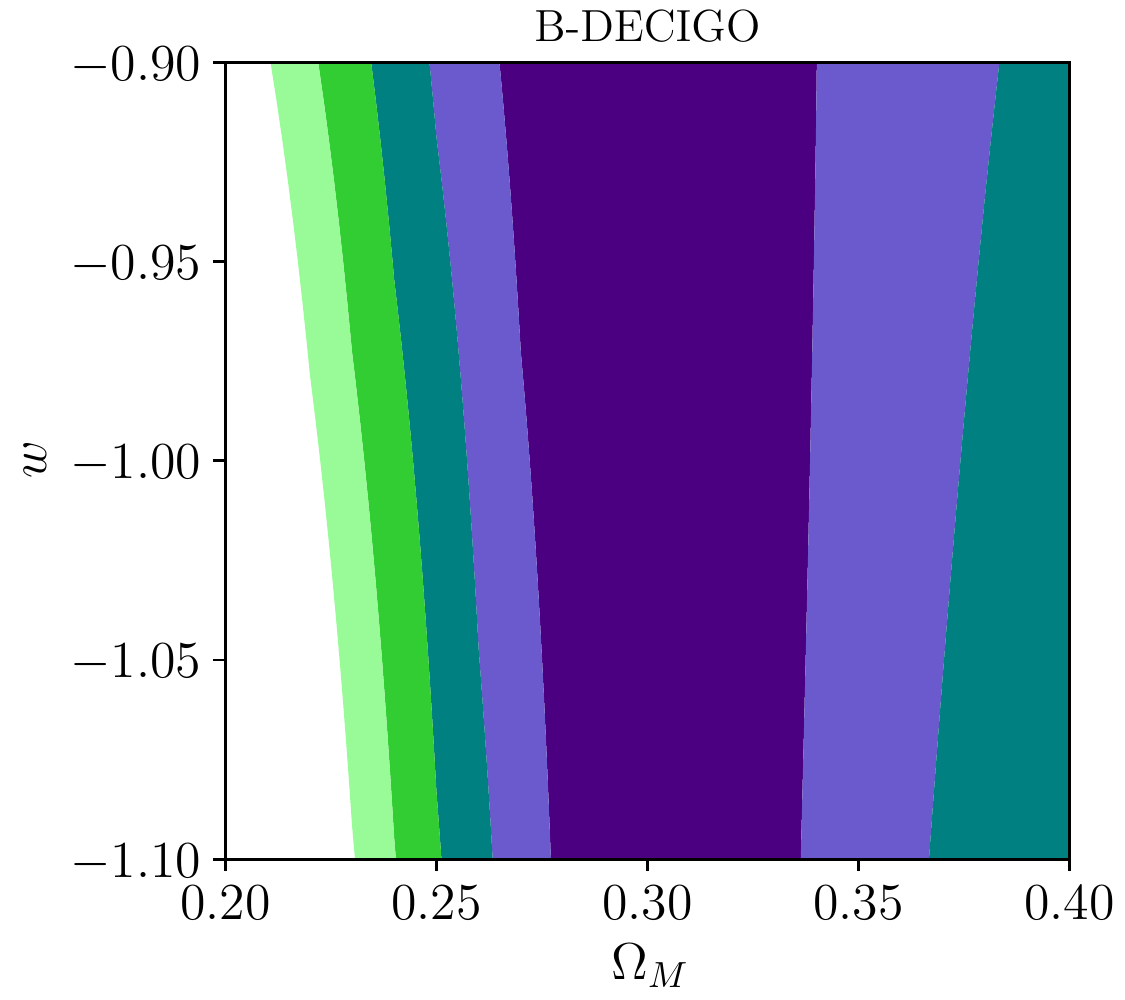}
  \caption{Constraints on $\Omega_M$ and $\omega$ from lensing statistics with $p_j=\ud\tau_j/\ud z_{\text{l},j}$ in Eq.~\eqref{eq-def-chis-1} for DECIGO (left panel) and B-DECIGO (right panel).}
  \label{fig-c-ls-zsu-zlk}
\end{figure*}

With these modified source and lens redshifts, one can still perform the lensing statistics and time delay methods to constrain either $(\Omega_M,\omega)$ or $(\Omega_M,h)$.
With lensing statistics, one can set either $p_j=\tau_j$ or $p_j=\ud\tau_j/\ud z_{\text{l},j}$ in Eq.~\eqref{eq-def-chis-1} as in the previous subsection.
Since $p_j=\tau_j$ did not lead to very tight constraints in the previous subsection, now, we simply show the bounds in the case of $p_j=\ud\tau_j/\ud z_{\text{l},j}$ in Fig.~\ref{fig-c-ls-zsu-zlk}.
Comparing this figure with Fig.~\ref{fig-c-ls-zlk} reveals that the bounds are worse, but DECIGO still constrains $\Omega_M$ a lot, i.e., $0.288\lesssim\Omega_M\lesssim0.314$.
As a final remark, one may notice that Figs.~\ref{fig-c-ls-zlk} and \ref{fig-c-ls-zsu-zlk} look very similar.
This is because although the redshifts of the sources and lenses are different for drawing the two figures, the probability function Eq.~\eqref{eq-def-p} is dominated by the contribution from the single-image events, whose redshift distribution does not change a lot (please compare the two lower  panels in Figs.~\ref{fig-rdec} and \ref{fig-rdec-u} for DECIGO, and the two lower panels in Figs.~\ref{fig-rbdec} and \ref{fig-rbdec-u}).

\begin{figure*}
  \centering
  \includegraphics[width=0.45\textwidth]{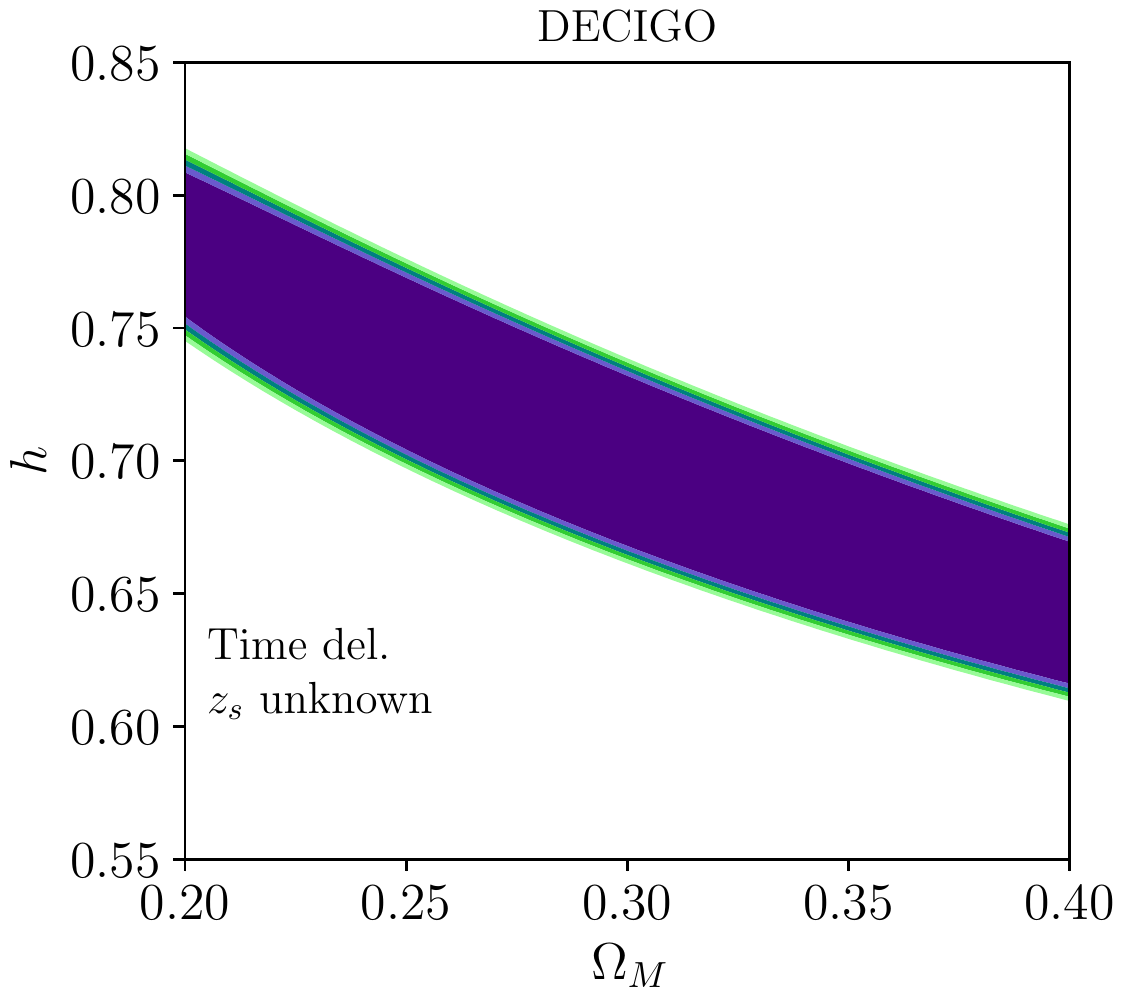}
  % the bounds on h: [0.623,0.777]
  \includegraphics[width=0.45\textwidth]{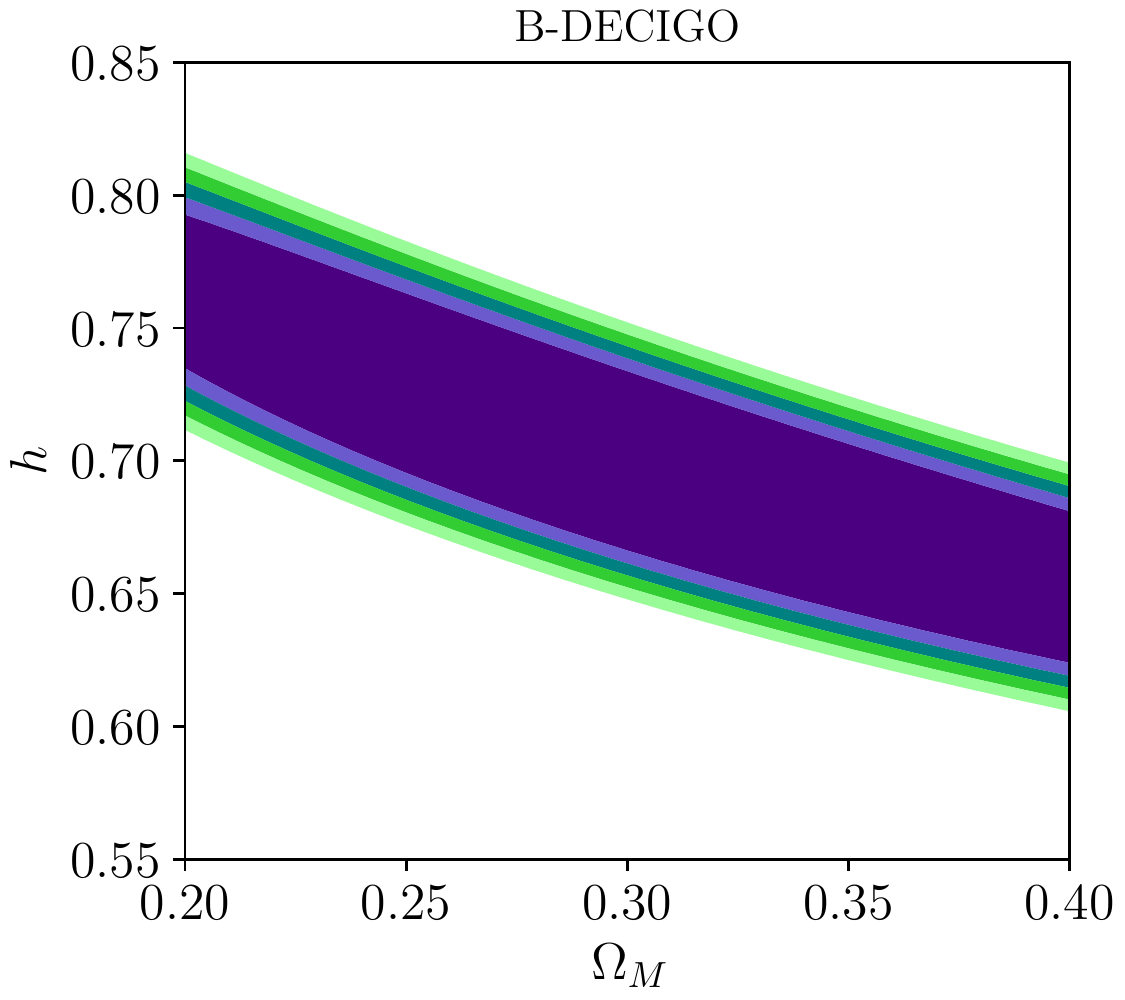}
  % the bounds on h: [0.596,0.804]
  \caption{Constraints on $\Omega_M$ and $h$ from time delay method for DECIGO (left panel) and B-DECIGO (right panel).}
  \label{fig-c-td-h0-u}
\end{figure*}

Now, consider the time delay method.
It does not provide better constraints on $\Omega_M$ and $w$, so we do not draw any figure here.
Instead, we simply present Fig.~\ref{fig-c-td-h0-u}, which displays the constraints on $\Omega_M$ and $h$.
The constraints become worse, and at $\Omega_M=0.3$, one obtains $0.623\lesssim h\lesssim0.777$ by DECIGO and $0.596\lesssim h\lesssim0.804$ by B-DECIGO.
Comparing with Table~1 in \citet{Sereno:2011ty}, the bounds on $h$ from DECIGO are at least half of those from LISA, and B-DECIGO's constraints are still stronger than LISA's but worse than DECIGO's.

\section{Conclusion}
\label{sec-con}

In this work, we make use of the predicted lensing rates of the future space-borne detectors DECIGO and B-DECIGO to constrain the cosmological parameters $\Omega_M$, $w$ and $H_0$.
Since they are mostly sensitive in the deci-hertz range, there are a lot of binary systems with small masses that can be the targets of the two detectors.
Therefore, it is predicted that each of them can detect tens to a few hundreds of multi-image events in 4 years' running \citep{Piorkowska-Kurpas:2020mst}.
Although nearly all of these events are binary black hole mergers, the constraints on cosmology model can still be obtained with the lensing statistics and time delay methods.
Both methods can put bounds on $\Omega_M$ and $w$ by fixing $h=0.7$, and the lensing statistics with $p_j=\ud\tau_j/\ud z_{\text{l},j}$ in Eq.~\eqref{eq-def-p} gives the best results.
It turns out that assuming the double compact objects formed via the standard evolutionary scenario with the low-end metallicity, the best constraints on $\Omega_M$ are obtained with DECIGO: $|\Delta\Omega_M|/\Omega_M\sim4\%$  at the $5\sigma$ level.
The constraints obtained using B-DECIGO are worse.
Unfortunately, one cannot bound $w$, at least in the range $w\in[-1.1,-0.9]$ we have chosen.
The time delay method can also be utilized to constrain $H_0$ and $\Omega_M$ at $w=-1$.
We found out that DECIGO may have $|\delta h|/h\sim3\%-11\%$, and B-DECIGO may give $|\delta h/h|\sim8\%-15\%$ at $5\sigma$ level.
These results are better than what were reported in \citet{Sereno:2011ty}.
By increasing the accuracy of measuring $d_L$ and $D_{\Delta t}$, all of the bounds are expected to be tighten further.

%%%%%%%%%%%%%%%%%%%%%%%%%%%%%%%%%%%%%%%%%%%%%%%%%%

\section*{Acknowledgements}

We are grateful for Hai Yu's help.
This work was supported by the National Natural Science Foundation of China under Grants No.~11633001, No.~11673008, No.~11922303, and No.~11920101003 and the Strategic Priority Research Program of the Chinese Academy of Sciences, Grant No. XDB23000000 and the Fundamental Research Funds for the Central Universities (No.~2042020kf1066).
SH was supported by Project funded by China Postdoctoral Science Foundation (No.~2020M672400).

\section*{Data Availability}

The simulation data are available from the corresponding author upon reasonable request.

\bibliographystyle{mnras}
%\bibliography{refs}
%\bibliography{lensingCosmography_v3.bbl}
\bibliography{../../References/refs}

% Don't change these lines
\bsp	% typesetting comment
\label{lastpage}
\end{document}